\documentclass[longbib, preprint]{aastex701}

\newcommand\afrho[1][\theta]{\mbox{$A(#1)f\rho$}}
\newcommand\afr{\mbox{$Af\rho$}}

\received{June 17, 2026}
\revised{August 10, 2026}
\accepted{August 20, 2026}
\submitjournal{The Planetary Science Journal}

\begin{document}

\title{The Highly Unusual Behavior and Chemical Composition of Interstellar Comet 2I/Borisov}

\author[0009-0006-9319-5645]{Allison N. Bair}
\affiliation{Lowell Observatory, 1400 W Mars Hill Road, Flagstaff, AZ 86001, USA}
\email{bair@lowell.edu}

\author[0000-0003-1412-2511]{David G. Schleicher}
\affiliation{Lowell Observatory, 1400 W Mars Hill Road, Flagstaff, AZ 86001, USA}
\email{dgs@lowell.edu}

\author[0000-0003-1008-7499]{Theodore Kareta}
\affiliation{Lowell Observatory, 1400 W Mars Hill Road, Flagstaff, AZ 86001, USA}
\affiliation{Department of Astrophysics and Planetary 
Science, Villanova University, Villanova, PA, USA}
\email{tkareta@lowell.edu}

\author[0000-0003-2781-6897]{Matthew M. Knight}
\affiliation{Volgenau Department of Physics, United States Naval Academy, 572C Holloway Road, Annapolis, MD 21402, USA}
\email{knight@usna.edu}

\author[0000-0002-6702-7676]{Michael S. P. Kelley}
\affiliation{Department of Astronomy, University of Maryland, College Park, MD 20742, USA}
\email{msk@astro.umd.edu}

\begin{abstract}

We report on imaging, spectroscopy, and narrowband photometry of interstellar comet 2I/Borisov obtained at Lowell Observatory from 2019 September 13 to December 16. We measured chemical composition on four nights before perihelion, from October 4 to November 25 (2.466 to 2.028 au), with CN detections on all nights. Our measurements from October 27 represent the earliest reported detections of C$_2$ and C$_3$, where C$_2$-to-CN and C$_3$-to-CN log production rate ratios were $-$0.7 and $-$1.3, respectively. Both ratios increased by nearly 4$\times$ to $-$0.3 and $-$0.9 by November 25, an extremely large increase with heliocentric distance that is unprecedented in the Lowell Observatory comet database. We additionally observed OH and NH on November 25, with a derived H$_2$O production rate of 4.5$\times $10$^{26}$ mol s$^{-1}$ and very high log production rate ratios for CN-to-OH ($-$2.0) and NH-to-OH ($-$1.7). Monitoring from broadband $R$ images from September 13 to December 16 suggests dust ($A(0^{\circ})f\rho$) production peaked around October 19, 50 days before its December 8 perihelion passage. Narrowband images from October 4 and 27 reveal a nearly symmetric CN distribution in Borisov's coma, suggesting activity over its entire sunward hemisphere and possibly its entire nucleus. Our results, combined with those from others, reveal a chemical composition and behavior with heliocentric distance unlike any other comet we have observed.

\end{abstract}

\keywords{\uat{Comets}{280}, \uat{Comae}{271}, \uat{Interstellar objects}{52}, \uat{Comet volatiles}{2162}}

\section{Introduction} 

The ices stored in the nuclei of comets contain chemical remnants of the environment from which they formed \citep[e.g.,][]{Mumma2011}. Over the last several decades, investigations of ensemble properties of solar system comets have revealed there are several chemical compositional classes of comets \citep{AHearn1995, Cochran2012, Fink2009, Schleicher2016} with multiple lines of evidence tying compositional differences to primoridal conditions when and where these comets were formed \citep{AHearn1995, DelloRusso2007, Bair2025}. The discovery of 2I/Borisov (henceforth Borisov), the second confirmed interstellar object detected in our solar system and the first to exhibit cometary activity, presented the first opportunity to investigate the chemical composition of a comet from a different planetary system. It was first detected inbound on 2019 August 30 at 2.99 au from the Sun, and early compositional measurements proved most interesting with clear detections of CN but only upper limits for the carbon-chain species C$_2$ and C$_3$ \citep[e.g.,][]{Fitzsimmons2019, Opitom2019, Kareta2020}. These early observations placed Borisov in the carbon-chain depleted class of comets as defined by \citet{AHearn1995}. More specifically, it was in the strongly carbon-chain depleted compositional class (i.e., strongly depleted in both C$_2$ and C$_3$ with respect to CN and OH) of the Lowell Observatory narrowband photometry database, which includes compositional measurements of over 230 comets observed since 1976 and incorporates the data published in \citet{AHearn1995} (see \citealt{Bair2025}). 

Unlike other comets in the Lowell Observatory database, which show negligible to small changes in composition with heliocentric distance, the combined measurements from multiple studies suggest Borisov exhibited exceptional compositional changes as it approached and passed perihelion \citep{Lin2020, Bair2020, Aravind2021, Deam2026}. While early upper limits of C$_2$ and C$_3$ suggested Borisov was strongly depleted in these species with respect to CN during early October 2019, by the time it reached its perihelion distance of 2.007 au on 2019 December 8 it exhibited only moderate carbon-chain depletion. A further indication of its unusual composition, Borisov had among the highest NH-to-OH abundance ratios in the Lowell Observatory comet database \citep{Bair2020} and was found to exhibit high NH$_2$-to-CN as compared to other cometary spectra \citep{Deam2026}. It additionally had an unusually high ratio of CO with respect to both water and HCN \citep{Bodewits2020, Cordiner2020}, as well as other compositional anomalies both before and after perihelion.

Our observational campaign consisted of imaging, spectroscopy, and photometry obtained with Lowell Observatory telescopes near Flagstaff, AZ during the last four months of 2019. Since telescope time had already been allocated prior to Borisov's discovery, we obtained as much data as was viable within our already awarded observing blocks. As discussed further below, we prioritized compositional studies, though Borisov's faintness throughout the period limited these investigations. In fact, Borisov's peak total brightness\footnote{\url{https://www.minorplanetcenter.net/db_search/show_object?utf8=\%E2\%9C\%93&object_id=2I}} only reached an apparent magnitude of $\sim15,$ which is roughly the threshold magnitude at which we {\it begin} our normal narrowband observations.

Borisov's visibility from Flagstaff complicated our studies; although initially discovered in the north, it crossed the celestial equator on 2019 November 13 and moved rapidly south, effectively becoming unobservable for us by the end of the year. Furthermore, its solar elongation was under 40$^\circ$ at discovery and remained below 80$^\circ$ throughout the observability window, restricting how long it was up on any given night. The Lowell Discovery Telescope's (LDT) instrument cube allowed rapid switching between imaging and spectroscopy so, despite short observing windows on our LDT nights, we obtained contextual imaging in addition to spectroscopy. Details involving instrumentation, our observations, and data reductions are presented in Section 2, while the results and discussion of our imaging and production rates are in Section 3, including a discussion of how our measurements compare to and fit into the timeline of compositional measurements from others. We additionally discuss the nucleus properties that can be deduced from our images and production rates, and consider Borisov's characteristics within in the context of solar system comets. A summary and  concluding remarks are in Section 4.

\section{Observations and Reductions} 
\label{sec:imaging_obs}

\subsection{CCD Imaging Observations and Reductions}
\begin{figure}
    \centering
    \includegraphics[width=0.65\linewidth]{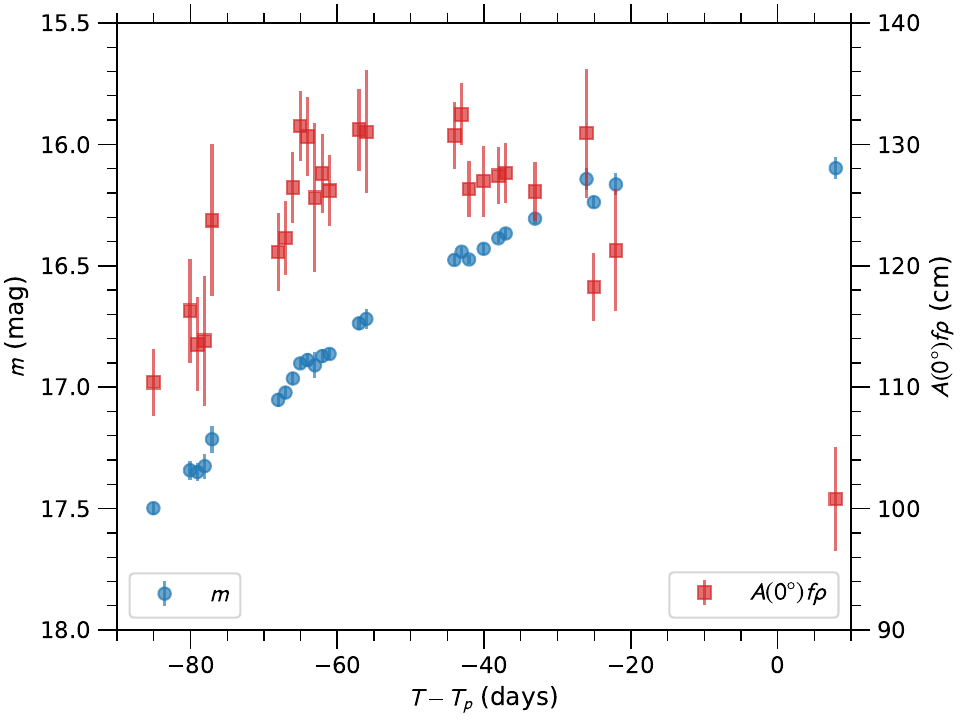}
    \caption{Broadband $r$ photometry (circles) of comet 2I/Borisov versus time from perihelion. The left vertical axis gives the magnitude, $m$, calibrated to the PS1-$r$ system.  The data are also converted to the cometary \afr{} parameter, and scaled by the phase function to a phase angle of 0\degr{} (squares). These values are given on the right vertical axis.\\ The photometry shown in this figure are available as the Data behind the Figure. 
    }
    \label{fig:robo-phot}
\end{figure}

We used Lowell Observatory's 31-in (0.8-m) telescope at Anderson Mesa to robotically monitor Borisov's brightness evolution and dust production while it was observable from Flagstaff, AZ, and successfully acquired short sets of broadband $R$ imaging on 27 nights from 2019 September 13 to December 16 (Table~\ref{t:imaging_circ}). Now decommissioned, the 31-in had an e2v CCD42-40 chip with 2K$\times$2K pixels and a pixel scale of 0.46$\arcsec$/pixel.
The telescope tracked at the comet's ephemeris rate during the exposures; exposure times were 180~s on all nights except 2019 September 14, when they were 300~s. As a result, background sources trailed 4\arcsec\ to 6\arcsec.  Instrumental calibrations (bias removal, flat fielding) were applied to the data. The positions and brightnesses of background sources were measured and used to derive the astrometric and photometric calibrations using \texttt{Calviacat} \citep{Kelley2019-calviacat}. For the photometry, the data were calibrated to the $r$-band of the PS1 photometric system, including a color correction, assuming a $g-r$ color of 0.53~mag \citep{Opitom2019}. Most frames matched 10 to 70 catalog sources, and calibration uncertainties were generally between 0.02 and 0.04 mag. To estimate the delivered image quality, we used the second-order moments of the background sources, assuming they are an elliptical shape. This assumption may not be true for the trailing direction, but is a reasonable approximation perpendicular to the trailing direction, which is mainly affected by seeing, optics, and tracking. The image quality estimate is 2.35 times the median of the background source semi-minor axes (3\arcsec\ to 4\arcsec). Photometry was measured in several aperture sizes. The typical 10,000~km constant radius aperture used in cometary astronomy was initially too small (4\arcsec\ to 7\arcsec) for the image quality; a 20,000~km aperture was more appropriate and ultimately used for this work. The resulting photometry is presented in magnitudes and as the cometary \afr{} parameter in Fig.~\ref{fig:robo-phot}. We use the quantity $A({\theta})f\rho$ as a proxy for dust production \citep{AHearn1984}; it is the product of the dust albedo ($A$) at the given phase angle ($\theta$), the filling factor for the aperture ($f$), and the projected aperture radius, $\rho$. We apply a first-order phase angle adjustment to the $A({\theta})f\rho$ values using the Schleicher-Marcus composite phase curve\footnote{\href{https://asteroid.lowell.edu/comet/dustphase/}{https://asteroid.lowell.edu/comet/dustphase/}} and normalize to 0$^\circ$ phase, i.e. $A(0^{\circ})f\rho$ (see \citealt{Schleicher1998} and \citealt{Schleicher2011}).

\begin{deluxetable}{lccccrccccccc}  
\tabletypesize{\scriptsize}
\tablecolumns{11}
\tablewidth{0pt} 
\setlength{\tabcolsep}{0.05in}
\tablecaption{Imaging Observations and Geometric Parameters.\tablenotemark{\scriptsize{a}}}
\tablehead{   
  \colhead{Date (UT)}&
  \colhead{UT range}&
  \colhead{Tel.\tablenotemark{\scriptsize{b}}}&
  \colhead{Airmass}&
  \colhead{$\Delta$T\tablenotemark{\scriptsize{c}}}&
  \colhead{$r_\mathrm{H}$\tablenotemark{\scriptsize{d}}}&
  \colhead{$\Delta$\tablenotemark{\scriptsize{e}}}&
  \colhead{$\theta$\tablenotemark{\scriptsize{f}}}&
  \colhead{P.A.\tablenotemark{\scriptsize{g}}}&
  \colhead{$R/r$\tablenotemark{\scriptsize{h}}}&
  \colhead{CN}&
  \colhead{VR}&
  \colhead{Conditions\tablenotemark{\scriptsize{i}}} \\
  \cline{10-12}
  \colhead{}&
  \colhead{}&
  \colhead{}&
  \colhead{}&
  \colhead{(day)}&
  \colhead{(AU)}&
  \colhead{(AU)}&
  \colhead{($^\circ$)}&
  \colhead{($^\circ$)}&
  \multicolumn{3}{c}{(\# $\times$ exposure time [s])}&
  \colhead{}
}
\startdata
2019 Sep 13 & 12:08-12:14 & 31in & 1.92 & $-$86.045 & 2.762 & 3.400 & 14.6 & 118.5 & 2$\times$180 & --- & --- & Spectro  [very smoky]\\
2019 Sep 14 & 11:32-12:18 & 31in & 2.13 & $-$85.056 & 2.747 & 3.378 & 14.8 & 118.2 & 9$\times$300 & --- & --- & Cloudy  [smoky]\\
2019 Sep 19 & 11:46-11:52 & 31in & 2.09 & $-$80.060 & 2.674 & 3.268 & 15.7 & 117.0 & 2$\times$180 & --- & --- & Clear\\
2019 Sep 20 & 11:43-11:50 & 31in & 2.10 & $-$79.062 & 2.660 & 3.246 & 15.9 & 116.8 & 2$\times$180 & --- & --- & Clear\\
2019 Sep 21 & 11:38-11:44 & 31in & 2.17 & $-$78.066 & 2.645 & 3.225 & 16.1 & 116.5 & 2$\times$180 & --- & --- & Clear\\
2019 Sep 22 & 11:38-11:45 & 31in & 2.14 & $-$77.065 & 2.631 & 3.203 & 16.3 & 116.3 & 2$\times$180 & --- & --- & Cloudy \\
2019 Oct 1 & 11:59-12:06 & 31in & 1.77 & $-$68.051 & 2.508 & 3.010 & 18.1 & 114.5 & 2$\times$180 & --- & --- & Clear \\
2019 Oct 2 & 11:58-12:04 & 31in & 1.79 & $-$67.052 & 2.495 & 2.990 & 18.3 & 114.3 & 2$\times$180 & --- & --- & Clear \\
2019 Oct 3 & 12:06-12:12 & 31in & 1.70 & $-$66.046 & 2.482 & 2.969 & 18.5 & 114.1 & 2$\times$180 & --- & --- & Clear \\
2019 Oct 4&11:55-12:02 & LDT & 1.69 & $-$65.054 & 2.470 & 2.948 & 18.7 & 114.0&1$\times$60&1$\times$300&---& Clear\\
2019 Oct 4&12:35-12:42& LDT & 1.33 & $-$65.026 & 2.469 & 2.947 & 18.7 & 114.0 &1$\times$60&1$\times$300&---& Clear [nautical twilight]\\
2019 Oct 5 & 11:56-12:03 & 31in & 1.77 & $-$64.053 & 2.457 & 2.927 & 18.9 & 113.8 & 2$\times$180 & --- & --- & Clear \\
2019 Oct 5&12:32-12:42& LDT & 1.48 & $-$64.027 & 2.456 & 2.927 & 19.0 & 113.8 &1$\times$30&3$\times$120&1$\times$30&Clear [nautical twilight]\\
2019 Oct 6 & 11:58-12:04 & 31in & 1.73 & $-$63.052 & 2.444 & 2.907 & 19.1 & 113.6 & 2$\times$180 & --- & --- & Clear \\
2019 Oct 7 & 11:54-12:00 & 31in & 1.77 & $-$62.055 & 2.432 & 2.886 & 19.3 & 113.4 & 2$\times$180 & --- & --- & Clear \\
2019 Oct 8 & 11:56-12:02 & 31in & 1.73 & $-$61.053 & 2.419 & 2.866 & 19.6 & 113.3 & 2$\times$180 & --- & --- & Clear \\
2019 Oct 12 & 12:15-12:21 & 31in & 1.56 & $-$57.040 & 2.371 & 2.785 & 20.4 & 112.6 & 2$\times$180 & --- & --- & Clear \\
2019 Oct 13 & 12:08-12:14 & 31in & 1.60 & $-$56.045 & 2.360 & 2.766 & 20.6 & 112.5 & 2$\times$180 & --- & --- & Clear \\
2019 Oct 25 & 12:02-12:08 & 31in & 1.59 & $-$44.049 & 2.232 & 2.540 & 22.9 & 110.9 & 2$\times$180 & --- & --- & Clear \\
2019 Oct 26 & 12:04-12:10 & 31in & 1.57 & $-$43.048 & 2.222 & 2.523 & 23.1 & 110.8 & 2$\times$180 & --- & --- & Clear \\
2019 Oct 27&10:54-11:15& LDT & 2.18 & $-$42.091 & 2.213 & 2.506 & 23.3 & 110.7 &3$\times$30&5$\times$180&---&Clear\\
2019 Oct 27 & 12:05-12:11 & 31in & 1.56 & $-$42.047 & 2.213 & 2.505 & 23.3 & 110.7 & 2$\times$180 & --- & --- & Cloudy \\
2019 Oct 27&12:55-13:06& LDT & 1.31 & $-$42.011 & 2.213 & 2.505 & 23.3 & 110.7 &1$\times$30&3$\times$180&---&Clear [nautical twilight]\\
2019 Oct 29 & 12:06-12:12 & 31in & 1.55 & $-$40.047 & 2.195 & 2.471 & 23.6 & 110.5 & 2$\times$180 & --- & --- & Clear \\
2019 Oct 30 & 12:00-12:06 & 31in & 1.59 & $-$39.051 & 2.186 & 2.454 & 23.8 & 110.4 & 2$\times$180 & --- & --- & Clear \\
2019 Oct 31 & 12:04-12:10 & 31in & 1.56 & $-$38.048 & 2.177 & 2.437 & 24.0 & 110.3 & 2$\times$180 & --- & --- & Clear \\
2019 Nov 1 & 12:07-12:13 & 31in & 1.54 & $-$37.046 & 2.169 & 2.421 & 24.2 & 110.2 & 2$\times$180 & --- & --- & Clear \\
2019 Nov 5 & 11:59-12:05 & 31in & 1.59 & $-$33.051 & 2.137 & 2.357 & 24.9 & 109.8 & 2$\times$180 & --- & --- & Cloudy\\
2019 Nov 12 & 12:21-12:27 & 31in & 1.49 & $-$26.036 & 2.088 & 2.254 & 26.0 & 109.4 & 2$\times$180 & --- & --- & Partial \\
2019 Nov 13 & 12:25-12:31 & 31in & 1.48 & $-$25.033 & 2.082 & 2.241 & 26.1 & 109.3 & 2$\times$180 & --- & --- & Cloudy\\
2019 Nov 16 & 12:27-12:33 & 31in & 1.48 & $-$22.032 & 2.066 & 2.202 & 26.6 & 109.2 & 2$\times$180 & --- & --- & Clear \\
2019 Dec 16 & 11:28-11:44 & 31in & 2.41 & \phantom{1}$+$7.931 & 2.014 & 1.957 & 28.7 & 110.7 & 2$\times$180 & 2$\times$300 & --- & Clear \\
\hline
\enddata
\vspace{-1mm}
\tablenotetext{a} {All parameters are given for the midpoint of the UT range.}\vspace{-3mm}
\tablenotetext{b} {Telescope used: 31in = 31 inch (0.8 m) telescope, LDT = Lowell Discovery Telescope.}\vspace{-3mm}
\tablenotetext{c} {Time from perihelion.}\vspace{-3mm}
\tablenotetext{d} {Heliocentric distance.}\vspace{-3mm}
\tablenotetext{e} {Geocentric distance.}\vspace{-3mm}
\tablenotetext{f} {Solar phase angle.}\vspace{-3mm}
\tablenotetext{g} {Position angle of the Sun measured from north through east.}\vspace{-3mm}
\tablenotetext{h} {$R$ used for all 31in nights, $r$ used for all LDT nights}\vspace{-3mm}
\tablenotetext{i} {31in conditions from Brian Skiff's archive of Flagstaff nighttime cloudiness, accessed 2026 June 4 ({http://www2.lowell.edu/cloudiness\_data/clouds.html})
``Partial''  means at least 3 consecutive cloud-free hours during the night, ``Spectro'' means less than about 1 mag extinction throughout the night.}
\label{t:imaging_circ}
\end{deluxetable}

Unfortunately, the $R$-band signal-to-noise ratio only afforded a limited assessment of the morphology. We additionally attempted imaging using the CN narrowband filter \citep{Farnham2000} with the 31-in telescope on two nights (November 8\footnote{This night is not listed in Table~\ref{t:imaging_circ} because conditions were too poor to yield usable photometry.} and December 16), but did not detect it through this filter on either night. Given these deficiencies, we do not discuss this aspect further for the 31-in data. 

Unsurprisingly, we had considerably better imaging results for the morphology when using the 4.3-m LDT, though we continued to be constrained by Borisov's faintness and the decreasing observing window as it moved south. We successfully imaged Borisov using the Large Monolithic Imager \citep[LMI;][]{Massey2013} on three nights: 2019 October 4, 5, and 27 (see Table~\ref{t:imaging_circ}). As noted earlier, our Borisov observations were acquired within blocks already allocated for other projects, and we prioritized measuring its composition spectroscopically. Thus, our imaging was brief each night: we obtained one SDSS-$r'$ (henceforth $r$) and one CN image both before and after obtaining spectra on October 4, a single set of VR, $r$, and CN after spectroscopy on October 5, and sets of $r$ and CN bracketing the spectroscopy on October 27. The sky was too bright during our post-spectroscopy sets and they did not prove useful for further analysis.

LMI has an e2v CCD with 6K$\times$6K pixels covering a field of view 12.3{\arcmin} on a side. Images were binned $2{\times}2$ yielding a pixel scale of 0.24\arcsec/pixel. All images were acquired at the comet's rate of motion, and small dithers were applied between consecutive frames in the same filter. We followed our standard reduction process \citep[e.g., ][]{Knight2021}, removing bias and performing flat-fielding with twilight flats. When multiple images were acquired in the same filter during a set, we combined them by subtracting a sky value taken near the edges of the frame, aligning on the centroid, and median combining the images. Owing to time constraints, we did not acquire observations necessary to remove continuum and absolutely calibrate the CN images to ``pure'' gas images. Thus, the CN images were ``contaminated'' with dust, the effect of which will be explored in Section~\ref{sec:imaging_results}.

All images were processed to increase the contrast of faint structures in the coma using standard image enhancement routines \citep{Schleicher2004, Samarasinha2014}. As usual \citep[e.g.,][]{Bair2018}, we favored azimuthal median subtraction since it is benign and is not reliant on parameter choices, although we also processed the images using other techniques such as division by a 1/$\rho$ profile ($\rho$ is the projected radial distance from the nucleus) and Laplace filtering since features can sometimes be more easily discerned in different methods of processing. Our interpretation of the morphology was consistent across enhancement techniques.

\subsection{Photometer Observations and Reductions}
\label{sec:phot_reductions}
Our photometry measurements were taken using an EMI 6256 photomultiplier tube and pulse-counting system on the Hall 42-in (1.1-m) telescope at Lowell Observatory in Flagstaff, AZ. The HB narrowband comet filters \citep{Farnham2000} were used to isolate gas emission bands in OH, NH, CN, C$_3$, and C$_2$ as well as continuum locations in the UV, Blue, and Green at 3448~\AA, 4450~\AA, and 5260~\AA, respectively. Three complete and two partial sets of traditional narrowband photoelectric photometry were obtained, with between one and four gas species detected during each attempt; ultimately, between the multiple sets, all five of the gas species were measured on our single night of photometry on 2019 November 25, approximately two weeks before Borisov reached perihelion. While several additional nights of photometry were planned, the weather was cooperative for only this single night; the observing circumstances are given in Table~\ref{t:phot_circ}.

\begin{deluxetable}{llrlcccccccc}  
\tabletypesize{\scriptsize}
\tablecolumns{12}
\tablewidth{0pt} 
\setlength{\tabcolsep}{0.05in}
\tablecaption{Narrowband Photometry Observing Circumstances and Fluorescence Efficiencies.\,\tablenotemark{a}}
\tablehead{   
  \multicolumn{3}{c}{UT Date}&
  \colhead{$\Delta$T\tablenotemark{b}}&
  \colhead{$r_\mathrm{H}$\tablenotemark{c}}&
  \colhead{$\Delta$\tablenotemark{d}}&
  \colhead{$\theta$\tablenotemark{e}}&
  \colhead{Phase Adj.\,\tablenotemark{f}}&
  \colhead{$\dot{r}_\mathrm{H}$\tablenotemark{g}}&
  \multicolumn{3}{c}{log $L/N$ (erg s$^{-1}$ molecule$^{-1}$)}\\
  \cline{10-12}
  \colhead{}&
  \colhead{}&
  \colhead{}&
  \colhead{(day)}&
  \colhead{(au)}&
  \colhead{(au)}&
  \colhead{($^\circ$)}&
  \colhead{}&
  \colhead{(km s$^{-1}$)}&
  \colhead{OH}&
  \colhead{NH}&
  \colhead{CN}
}
\startdata
2019&Nov&25.50&\phantom{0}$-$13.1&2.028&2.099&27.6&2.311&\phantom{0}$-$5.5&$-$15.412&$-$13.830&$-$13.169\\
\hline
\enddata
\vspace{-1mm}
\tablenotetext{a} {All parameters are given for the midpoint of the night's observations.}\vspace{-3mm}
\tablenotetext{b} {Time from perihelion.}\vspace{-3mm}
\tablenotetext{c} {Heliocentric distance.}\vspace{-3mm}
\tablenotetext{d} {Geocentric distance.}\vspace{-3mm}
\tablenotetext{e} {Solar phase angle.}\vspace{-3mm}
\tablenotetext{f} {Adjustment to 0$^\circ$ solar phase angle to log($A(0^{\circ})f\rho$) values based on assumed phase function (see text).}\vspace{-3mm}
\tablenotetext{g} {Heliocentric velocity.}
\label{t:phot_circ}
\end{deluxetable}

We used our standard reduction methods, detailed in \citet{AHearn1995}, while information on how we derived fluxes is in \citet{Farnham2000}. The resulting emission band flux values for each gas species following continuum subtraction, along with the continuum fluxes, are listed in Table~\ref{t:phot_flux}. Fluxes for the gas species were converted into column abundances by applying the appropriate fluorescence efficiencies ($L/N$). The $L/N$ values for C$_2$ and C$_3$ are given in Table~2 of \cite{AHearn1995} and scale as $r_\mathrm{H}^{-2}$ where $r_\mathrm{H}$ is heliocentric distance, while for OH the values vary with heliocentric velocity ($\dot{r}_\mathrm{H}$) due to the Swings effect \citep{Schleicher1988}, and for CN and NH they vary due to both the Swings effect and heliocentric distance \citep{Schleicher2010, Meier1998}; the resulting L/N values for OH, CN, and NH are listed in Table \ref{t:phot_flux}. We applied a standard Haser model \citep{Haser1957} to extrapolate total coma abundances from the column abundances, with gas production rates computed by dividing by the assumed daughter lifetimes. The Haser parent and daugher scalelengths, in addition to daughter lifetimes, are tabluated in \citet{AHearn1995}, with all values assumed to scale as $r_\mathrm{H}^{2}$. The final column abundances (log~$M$($\rho$)) are listed in Table~\ref{t:phot_flux} while the derived production rates (log~$Q$) are in Table \ref{t:phot_rates}. Since OH has only one parent molecule, H$_2$O, we converted our OH Haser production rates to vectorial equivalent water production rates using the empirical relation derived by \citet{Cochran1993} and discussed in \citet{Schleicher1998}. The resulting derived water production value for each observational set is in the last column of Table \ref{t:phot_rates}. We use the quantity $A({\theta})f\rho$, a proxy for dust production as defined in Section 2.1, for each continuum filter. This value is listed in Table \ref{t:phot_rates}, while the first-order phase angle adjustment to normalize to 0$^\circ$ phase for our $A({\theta})f\rho$ values (see Section 2.1) is listed in Table \ref{t:phot_circ}.

\begin{deluxetable}{lccccccccc@{\extracolsep{4pt}}ccc@{\extracolsep{4pt}}ccccc}  
\tabletypesize{\scriptsize}
\tablecolumns{18}
\tablewidth{0pt} 
\setlength{\tabcolsep}{0.03in}
\tablecaption{Photometric Fluxes and Aperture Abundances.}
\tablehead{   
  \multicolumn{3}{c}{UT Date}&
  \multicolumn{2}{c}{Aperture}&
  \multicolumn{5}{c}{log Emission Band Flux\tablenotemark{a}}&
  \multicolumn{3}{c}{log Continuum Flux\tablenotemark{a}}&
  \multicolumn{5}{c}{ log $M$($\rho$)\tablenotemark{a} }\\
  \cline{4-5}
  \colhead{}&
  \colhead{}&
  \colhead{}&
  \colhead{Size}&
  \colhead{log $\rho$}&
  \multicolumn{5}{c}{(erg cm$^{-2}$ s$^{-1}$)} &
  \multicolumn{3}{c}{(erg cm$^{-2}$ s$^{-1}$ \AA$^{-1}$)} &
  \multicolumn{5}{c}{(molecule)}\\
  \cline{6-10}
  \cline{11-13}
  \cline{14-18}
  \colhead{}&
  \colhead{}&
  \colhead{}&
  \colhead{(arcsec)}&
  \colhead{(km)}&
  \colhead{OH}&
  \colhead{NH}&
  \colhead{CN}&
  \colhead{C$_3$}&
  \colhead{C$_2$}&
  \colhead{UV}&
  \colhead{Blue}&
  \colhead{Green}&
  \colhead{OH}&
  \colhead{NH}&
  \colhead{CN}&
  \colhead{C$_3$}&
  \colhead{C$_2$}
}
\startdata
2019&Nov&25.46&\phantom{0}77.8&4.77&...&...&$-$11.85&$-$12.10&$-$12.11&{\it und}&$-$14.92&$-$15.67&...&...&29.30&28.60&28.94\\
2019&Nov&25.49&\phantom{0}77.8&4.77&{\it und}&$-$13.01&$-$11.90&{\it und}&$-$12.46&$-$14.69&{\it und}&$-$14.48&{\it und}&28.77&29.25&{\it und}&28.59\\
2019&Nov&25.50&\phantom{0}77.8&4.77&{\it und}&$-$12.49&$-$11.83&$-$11.92&$-$12.29&$-$14.55&$-$14.83&$-$14.81&{\it und}&29.29&29.32&28.79&28.76\\
2019&Nov&25.51&\phantom{0}77.8&4.77&$-$11.84&$-$12.45&$-$11.83&{\it und}&$-$12.09&$-$15.53&$-$14.69&$-$14.66&31.63&29.33&29.32&{\it und}&28.97\\
2019&Nov&25.52&\phantom{0}77.8&4.77&&...&...&...&...&$-$15.53&...&...&&...&...&...&...\\
\hline
\enddata
\vspace{-1mm}
\tablenotetext{a} {{\it ``und''} stands for ``undefined''. For the gases, this means that the emission flux was measured but was less than zero following sky and continuum removal. For the continuum, this means the continuum flux was measured but following sky subtraction it was less than zero. ``...'' indicates that no measurements were attempted.}
\label{t:phot_flux}
\end{deluxetable}

\begin{deluxetable}{lllcccccccc@{\extracolsep{4pt}}ccc@{\extracolsep{4pt}}c}  
\tabletypesize{\scriptsize}
\tablecolumns{15}
\tablewidth{0pt} 
\setlength{\tabcolsep}{0.04in}
\tablecaption{Photometric and Spectroscopic Production Rates.}
\tablehead{   
  \multicolumn{3}{c}{UT Date}&
  \colhead{$\Delta$T}&
  \colhead{log $r_\mathrm{H}$}&
  \colhead{log $\rho$}&
  \multicolumn{5}{c}{log $Q$\tablenotemark{a,}\tablenotemark{b}\phantom{00}(molecules s$^{-1}$)}&
  \multicolumn{3}{c}{log $A$($\theta$)$f\rho$\tablenotemark{a,}\tablenotemark{b}\phantom{0}(cm)}&
  \colhead{log $Q$}\tablenotemark{a}\\
  \cline{7-11}
  \cline{12-14}
  \cline{15-15}
  \colhead{}&
  \colhead{}&
  \colhead{}&
  \colhead{(day)}&
  \colhead{(au)}&
  \colhead{(km)}&
  \colhead{OH}&
  \colhead{NH}&
  \colhead{CN}&
  \colhead{C$_3$}&
  \colhead{C$_2$}&
  \colhead{UV}&
  \colhead{Blue}&
  \colhead{Green}&
  \colhead{H$_2$O}
}
\startdata
\multicolumn{3}{l}{Spectroscopy}\\
2019&Oct& 4.51&\phantom{0}$-$65.0&0.392&...&...&...&24.13{\tiny\phantom{.}.02}&$<$24.2&$<$23.1&...&...&1.95{\tiny\phantom{.}.09}&...\\
2019&Oct& 5.52&\phantom{0}$-$64.0&0.390&...&...&...&23.98{\tiny\phantom{.}.05}&$<$24.3&$<$23.5&...&...&1.70{\tiny\phantom{.}.20}&...\\
\vspace{0.1in}
2019&Oct&27.50&\phantom{0}$-$42.0&0.345&...&...&...&24.39{\tiny\phantom{.}.02}&\phantom{1}23.1{\tiny\phantom{.}.1}&\phantom{1}23.7{\tiny\phantom{.}.1}&...&...&1.78{\tiny\phantom{.}.04}&...\\
\multicolumn{3}{l}{Photometry}\\
2019&Nov&25.46&\phantom{0}$-$13.1&0.307&4.77&...&...&24.63{\tiny\phantom{.}.04}&23.90{\tiny\phantom{.}.23}&24.47{\tiny\phantom{.}.17}&{\it und}&1.23{\tiny\phantom{.}.40}&0.51{\tiny\phantom{.}.94}&...\\
2019&Nov&25.49&\phantom{0}$-$13.1&0.307&4.77&{\it und}&24.47{\tiny\phantom{.}.69}&24.58{\tiny\phantom{.}.07}&{\it und}&24.12{\tiny\phantom{.}.67}&1.78{\tiny\phantom{.}.29}&{\it und}&1.69{\tiny\phantom{.}.17}&{\it und}\\
2019&Nov&25.50&\phantom{0}$-$13.1&0.307&4.77&{\it und}&25.00{\tiny\phantom{.}.16}&24.65{\tiny\phantom{.}.04}&24.08{\tiny\phantom{.}.16}&24.29{\tiny\phantom{.}.18}&1.93{\tiny\phantom{.}.21}&1.32{\tiny\phantom{.}.31}&1.36{\tiny\phantom{.}.30}&{\it und}\\
2019&Nov&25.51&\phantom{0}$-$13.0&0.307&4.77&27.12{\tiny\phantom{.}.09}&25.03{\tiny\phantom{.}.15}&24.65{\tiny\phantom{.}.03}&{\it und}&24.49{\tiny\phantom{.}.12}&0.94{\tiny\phantom{.}.81}&1.46{\tiny\phantom{.}.24}&1.52{\tiny\phantom{.}.23}&27.10\\
\vspace{0.05in}2019&Nov&25.52&\phantom{0}$-$13.0&0.307&4.77&26.73{\tiny\phantom{.}.09}&...&...&...&...&0.94{\tiny\phantom{.}.81}&...&...&26.71\\
\multicolumn{3}{l}{mean Nov 25.50\tablenotemark{c}}&\phantom{0}$-$13.1&0.307&4.77&26.67{\tiny\phantom{.}.22}&24.93{\tiny\phantom{.}.18}&24.63{\tiny\phantom{.}.02}&23.70{\tiny\phantom{.}.09}&24.37{\tiny\phantom{.}.09}&1.51{\tiny\phantom{.}{.20}}&1.22{\tiny\phantom{.}{.18}}&1.43{\tiny\phantom{.}{.11}}&26.65\\
\hline
\enddata
\vspace{-1mm}
\tablenotetext{a} {Production rates followed by the upper, i.e. the positive, uncertainty. Uncertainties are based on photon statistics and reflect the 1- $\sigma$ values derived from the propagation of the observational uncertainties. The ``+'' and ``$-$'' uncertainties are equal as percentages, but unequal in log-space; the ``$-$'' uncertainty values can be computed. Upper limits are $3\sigma$.}
\tablenotetext{b} {{\it ``und''} stands for ``undefined''. For the gases, this means that the emission flux was measured but was less than zero following sky and continuum removal. For the continuum, this means the continuum flux was measured but following sky subtraction it was less than zero. ``...'' indicates that no measurements were attempted.}\vspace{-1mm}
\tablenotetext{c}{Mean of the photometry measurements for each species on 2019 Nov 25. In the cases where a measurement went negative after continuum subtraction due to too low signal-to-noise, the value was included as a zero since a measurement was attempted}
\label{t:phot_rates}
\end{deluxetable}

\subsection{Spectral Observations and Reductions}

We obtained long-slit spectra of Borisov with LDT on the same three nights as used for imaging. LDT's DeVeny spectrograph \citep{Bida2014} is an e2v CCD42-10 deep depletion device having $2048{\times}512$ pixels. The slit length is 2.5{\arcmin} and the pixel scale in the spatial direction is 0.34\arcsec/pixel. On all nights we used the DV2 grating, which has 300 lines/mm and yields a spectral range of 3000 to 7400~\AA\ and a resolving power of $R=800$ at 5200~\AA. For science images the slit width was set to 3.0\arcsec\ and the slit was aligned at the parallactic angle. The telescope was tracked at the comet's rate during Borisov exposures and all observations appear unsmeared and properly executed.

Biases and dome flats were acquired at the start of the night on October 4 and 27 and immediately prior to using the DeVeny spectrograph on October 5. Directly before observing Borisov, two flux standard stars were observed at different airmasses along with one (October 4 and 5) or two (October 27) solar analogs for continuum removal. Immediately after observing Borisov, wavelength calibration lamps were observed at the comet's position. See Table~\ref{t:spectroscopy_circ} for an observing log.

\begin{deluxetable}{llcccc}  
\tabletypesize{\scriptsize}
\tablecolumns{11}
\tablewidth{0pt} 
\setlength{\tabcolsep}{0.05in}
\tablecaption{Spectroscopy observing log. Conditions and geometric parameters given in Table~\ref{t:imaging_circ}.\tablenotemark{\scriptsize{a}}}
\tablehead{  
  \colhead{Date (UT)}&
  \colhead{Object}&
  \colhead{UT range}&
  \colhead{Airmass}&
  \colhead{\# $\times$ exposure time [s]}&
  \colhead{Comment}
}
\startdata
2019 Oct 4\\
\hline
& PG0934+554 & 11:32-11:36 & 1.59 & 1$\times$30, 3$\times$60 & Flux standard\\
& HD 28099 & 11:40-11:44 & 1.06 & 1$\times$10, 3$\times$60 & Solar analog\\
& G191-B2B & 11:47-11:51 & 1.05 & 1$\times$10, 3$\times$60 & Flux standard\\
& 2I/Borisov & 12:06-12:28 & 1.63 & 1$\times$30, 4$\times$300 & Science target\\
& Hg+Cd+Ar lamps & 12:30-12:33 & 1.52 & 6$\times$15 & Comparison lamp\\
\hline
\\
\hline
2019 Oct 5\\
\hline
& PG0934+554 & 11:52-12:00 & 1.48 & 6$\times$60 & Flux standard\\
& HD 28099 & 12:02-12:04 & 1.09 & 1$\times$60, 3$\times$10 & Solar analog\\
& G191-B2B & 12:06-12:11 & 1.06 & 4$\times$60 & Flux standard\\
& 2I/Borisov & 12:16-12:27 & 1.59 & 1$\times$30, 2$\times$300 & Science target\\
& Hg+Cd+Ar lamps & 12:28-12:30 & 1.53 & 5$\times$15 & Comparison lamp\\
\hline
\\
\hline
2019 Oct 27\\
\hline
& PG0934+554 & 10:18-10:23 & 1.51 & 1$\times$10, 3$\times$60 & Flux standard\\
& HD 28099 & 10:28-10:30 & 1.08 & 5$\times$10 & Solar analog\\
& G191-B2B & 10:33-10:39 & 1.05 & 1$\times$10, 5$\times$60 & Flux standard\\
& Feige 34 & 10:43-10:50 & 1.82 & 1$\times$10, 5$\times$60 & Flux standard\\
& 2I/Borisov & 11:21-12:47 & 1.81 & 1$\times$30, 16$\times$300 & Science target\\
& Hg+Cd+Ar lamps & 12:49-12:50 & 1.34 & 3$\times$15 & Comparison lamp\\
\hline
\enddata
\vspace{-1mm}
\tablenotetext{a} {All parameters are given for the midpoint of the UT range.}
\label{t:spectroscopy_circ}
\end{deluxetable}

The spectra were reduced and extracted using the \textit{Pypeit} package \citep{2020JOSS....5.2308P} with settings specific to DeVeny. This package performs basic reduction steps (debiasing, flatfielding, deriving wavelength solutions, and subsequent extraction of spectra) automatically but with significant flexibility in terms of technique. Given that Borisov was clearly extended by several arcseconds in the frames even before basic calibration, multiple different approaches were attempted to perform background subtraction on the spectra. Cometary spectra are most extended at wavelengths with significant molecular emission, often significantly beyond what can be seen in a single exposure, and as such common methods for background subtraction (e.g., taking an average of most of the frame as a function of wavelength) can diminish or remove weak emission features easily. The most successful approach was to turn off the `2dmodel\_mask' and `local\_sky' settings within \textit{Pypeit}. The former allows for extended emission line sources (like comets and galaxies) to not have these areas of extended emission not rejected through the pipeline's extraction algorithm which otherwise would assume wavelengths with radically different PSFs were artifacts. The latter also aids in the extraction of emission line sources by forcing the pipeline to only consider information about the background from outside of the extraction region. Given that Borisov was the brightest object in the slit and identification of the target was easy in single exposures, the alternative background subtraction methods, such as specifying that no background was utilized within $\sim30-40$\arcsec\ of Borisov, produced relatively similar results (e.g., changes to our final derived production rates less than $1\sigma$).

The spectra were extracted with a range of apertures in a boxcar method. They were then analyzed to determine the optimal aperture for maximizing the signal-to-noise ratio of the CN feature. We used the same aperture for the other gas species, whose signal-to-noise ratios were too low to allow such analysis. Varying the extraction aperture size also gave a qualitative way to assess the reliability of other weaker gas features. Under most circumstances, the gases should be detected with higher confidence as the extraction aperture grows until a point at which the read noise from the detector (or contamination by other sources, negligible in this case) begins to matter significantly. Furthermore, the amount of enclosed flux should approximate a Haser profile \citep{Haser1957} when assessed at multiple apertures.

On October 4 and 27, which both had relatively good conditions, we extracted $5.65$\arcsec\ and $6.46$\arcsec\ along the slit's length (this is the boxcar radius mentioned above), though similar results were obtained if those apertures were changed by a pixel in radius ($\pm0.5\sigma$ or less in CN production rate). On October 5, where only two long exposures were obtained and conditions were less ideal, only $3.74$\arcsec\ were extracted along the slit. While in each of these cases more cometary signal must have been at larger radii, the read noise of the detector became noticeable and the significance of our gas detections dropped when extraction radii larger than these amounts were utilized. We inspected the cross-slit spatial profiles of the highest-quality October 27 data to get a sense of how much of the well-detected CN flux might be getting over-subtracted by our background subtraction procedure described above and found that $>99\%$ of the CN flux on the chip was within the region excluded from background estimation. After the extraction aperture was selected for each night, the same extraction procedures as were used for Borisov were applied to solar and flux calibrator stars (e.g., the same background subtraction and combination procedures were applied) observed that evening. We tested extracting our star spectra with both \textit{Pypeit}-determined optimal apertures and apertures that matched the size used for our comet spectra and again found minimal ($<<1\sigma$) differences between the two. All fluxes reported are from the latter approach, though again we stress that the difference between the two was neglible. All spectra had a theoretical extinction curve, derived from observations taken at Kitt Peak National Observatory, applied based on their airmass as recommended by the instrument manual, and were then combined in a per-target robust average. The validity of the extinction curve to each night of data was verified through comparison of flux calibration results utilizing standard stars observed at different airmasses. The combined Borisov spectra are strictly of the 300 second exposures on each night, as adding in additional shorter exposures did not meaningfully contribute to the signal-to-noise ratio of the extracted spectra primarily due to increased read-noise making the fainter features more challenging to discern than in the longer exposures. The cometary and solar analog spectra were flux calibrated using one of our standard star observations, and each night of calibrations was verified to be statistically identical regardless of which stars were utilized, adding confidence to our flux-calibration procedure.

The extracted cometary spectrum includes a continuum component composed of reflected light from dust onto which the emission features are super-imposed. Following \citet{Kareta2020}, we selected the wavelengths of light corresponding to the neutral or continuum filters in the HB filter set \citep{Farnham2000}, and then fit either a low-order polynomial or a linearly interpolated spline through each of these wavelength ranges or through the averages of the points within them (e.g., mimicking the average reflectivity measurement one might get from using those continuum filters). While the two approaches provided statistically identical results, we utilized a third-order polynomial fit in all final extractions for consistency. This approach was then used to derive $Af\rho$ measurements through the green continuum filter, though the precision of these spectra-derived measurements is significantly lower than our photometric $Af\rho$ values obtained through direct photometric imaging observations or with the photometer. 

\begin{figure}
\centering
	\includegraphics[width=0.70\columnwidth]{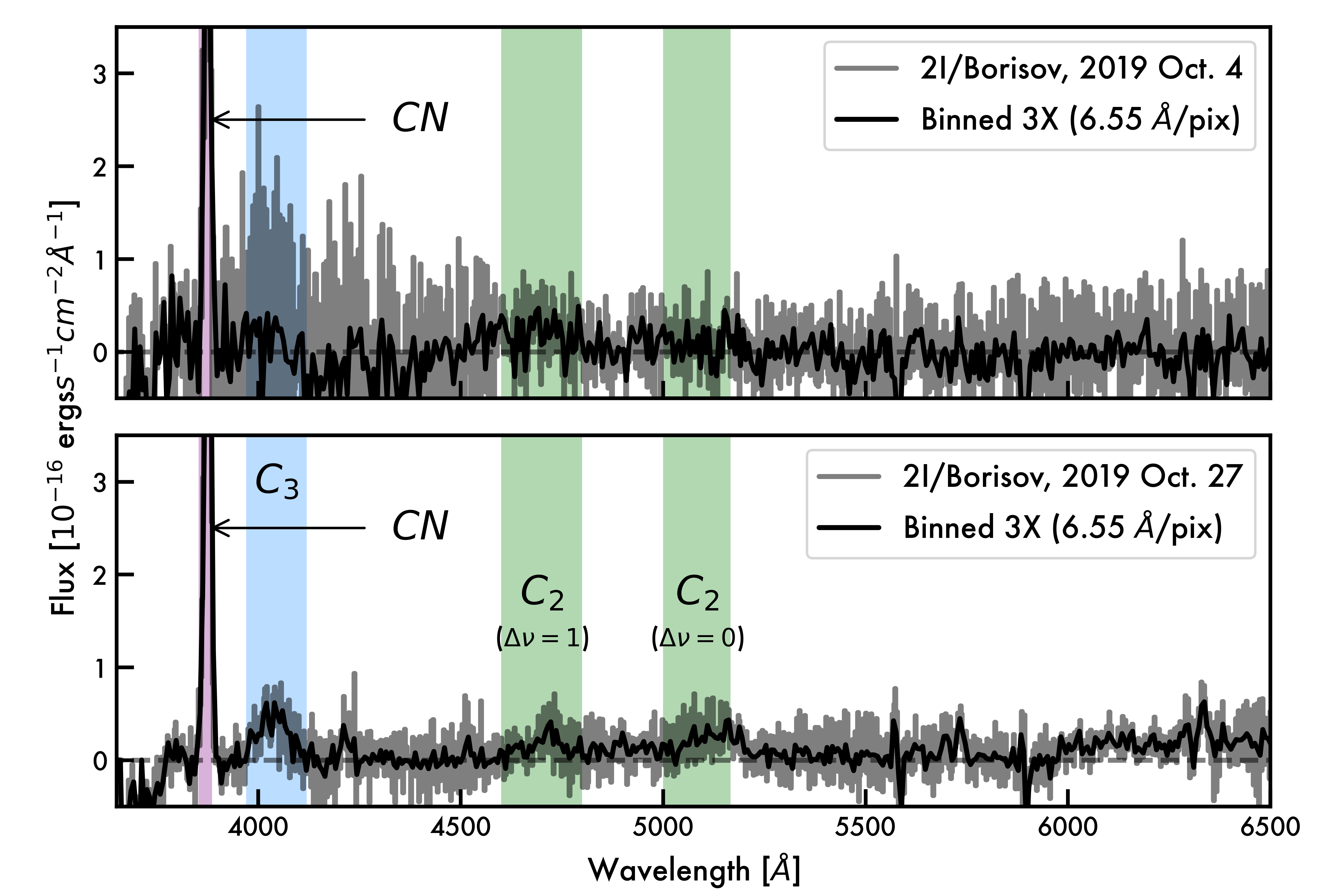}
    \caption{The flux-calibrated and continuum-removed spectrum of 2I/Borisov as measured on 2019 October 4 (top) and October 27 (bottom) with DeVeny on the LDT is shown. The unbinned spectrum on which all analyses were performed is shown in grey while a $3\times$ binned spectrum is shown in black. The wavelength ranges over which we extracted flux from CN, C$_3$, and C$_2$ emissions are shown in purple, blue, and green, respectively. All three species were detected on October 27, while only CN was detected in the earlier datasets.}
    \label{fig:oct27spec}
\end{figure}

We incorporated differences in our subtraction approaches' outcomes into our final error estimates on band fluxes, though they are typically minor. The band fluxes were then input, along with information about the comet's position and velocity, into the Haser model as implemented and validated in \citet{Kareta2020} with the scalelengths and lifetimes of \citet{AHearn1995}. While we address the applicability of Solar System comet-derived scalelengths to Borisov the next section, we employed them here to be consistent with other workers who reported production rates around the same time. This process was repeated for each species in the spectrum regardless of whether or not it was confidently detected or reported as an upper limit. The flux-calibrated and continuum-subtracted spectrum of Borisov on October 27th is shown and labeled in Figure \ref{fig:oct27spec}, and our derived production rates for each of the three nights are listed in Table~\ref{t:phot_rates}.

Searches for other features that might be detectable given the wavelength range in question --- namely, OH and NH --- were unsuccessful. We sought them both using the above approach (summing fluxes over their specific wavelength ranges) as well as detailed investigation of the 2D spectra themselves. The lack of a detection of OH, NH, and NH$_2$ is ultimately unsurprising given the faintness of Borisov, the moderate to large airmasses at which we observed, and the associated low signal-to-noise ratio. Furthermore, [OI] is not distinguishable from telluric oxygen at our low-to-moderate spectral resolution.

\section{Results and Discussion}

\subsection{Morphology}

\label{sec:imaging_results}

\begin{figure}
\centering
	\includegraphics[width=0.48\columnwidth]{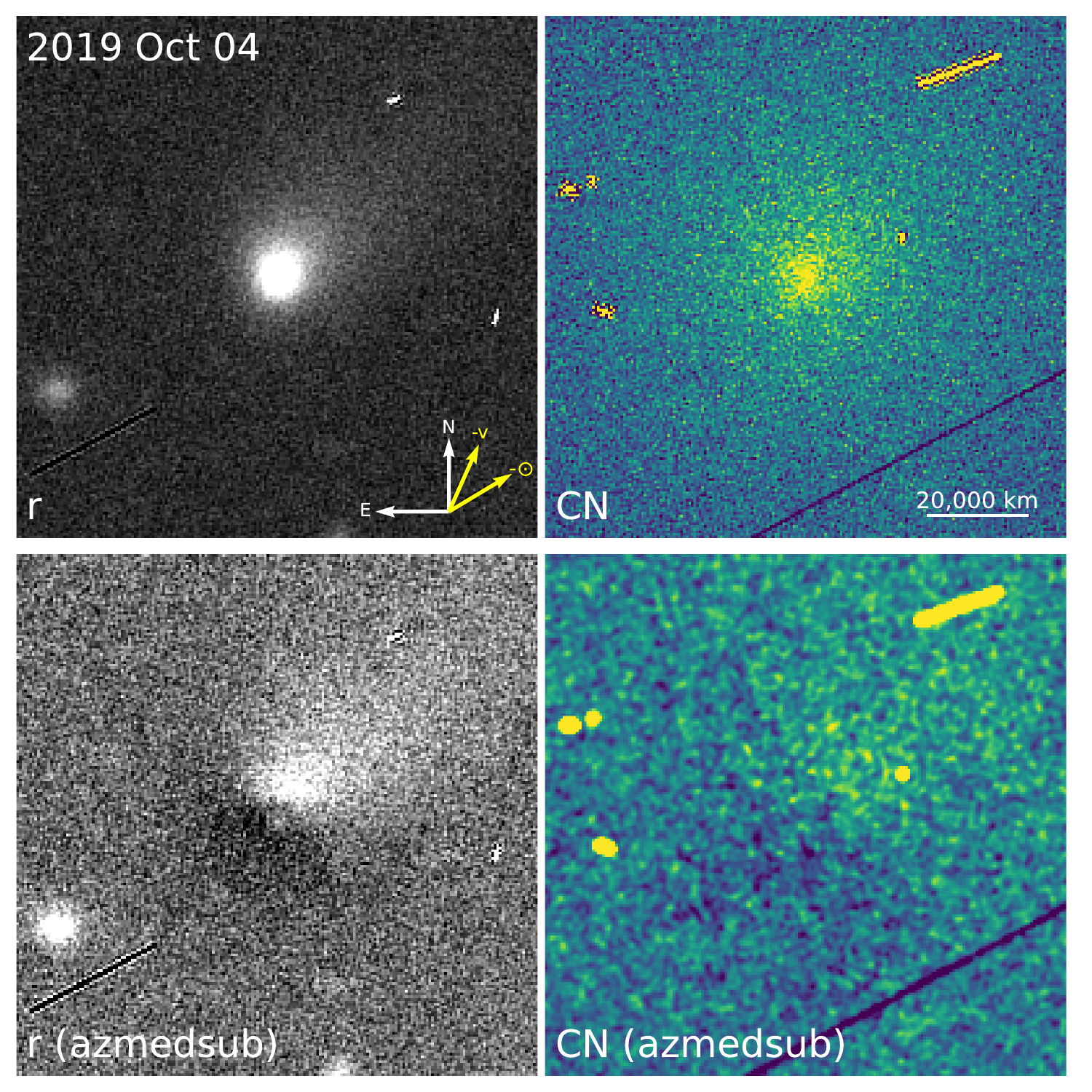}
	\includegraphics[width=0.48\columnwidth]{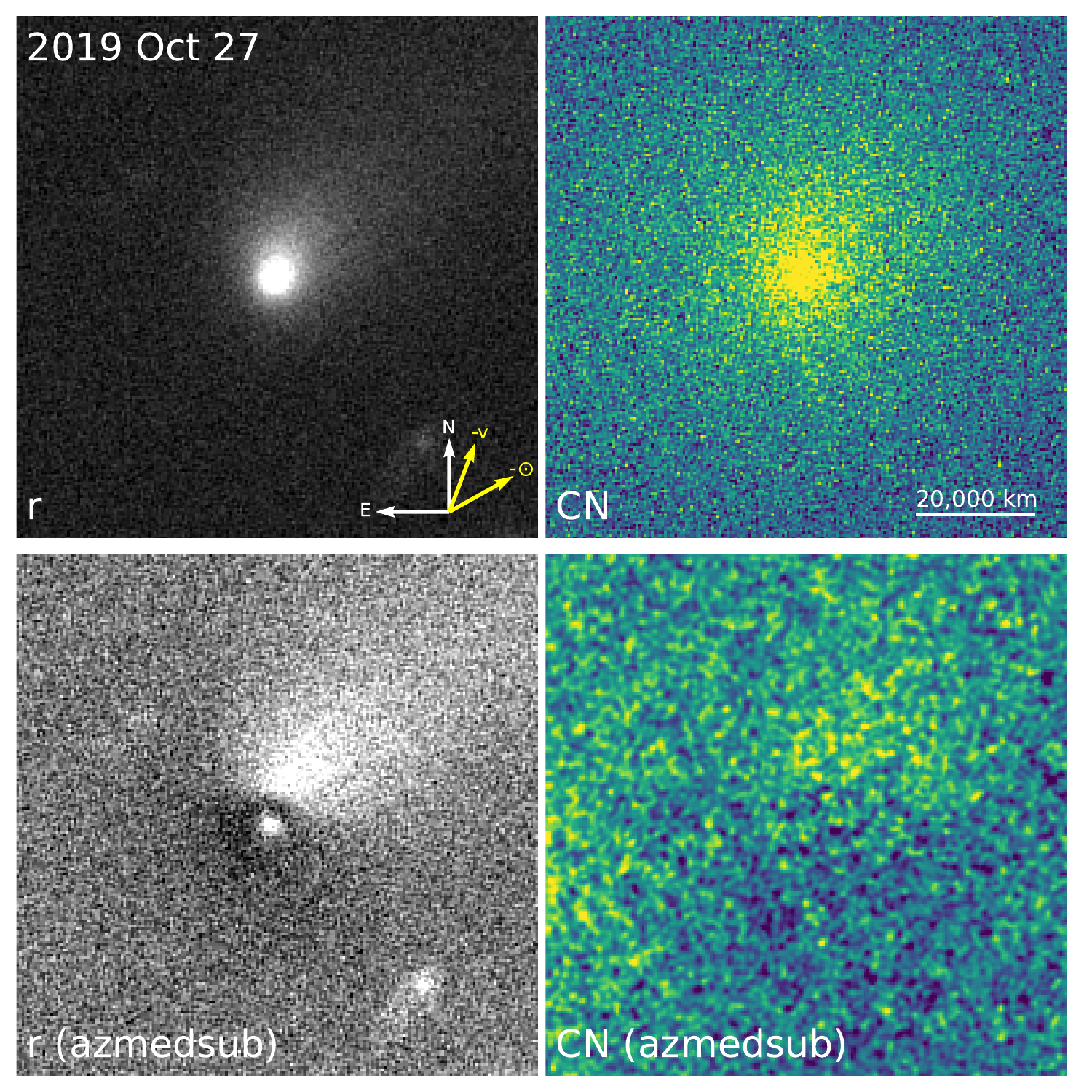}
    \caption{Images acquired on 2019 October 4 (left) and 27 (right) with LDT. The top row show stacks of $r$ and CN images, respectively. The bottom row shows the same images enhanced by subtraction of an azimuthal median profile. The enhanced CN images have been smoothed with a Gaussian profile to better reveal the faint underlying structure. A compass indicating north (N), east (E), the anti-sunward direction ($-\odot$), and the anti-velocity vector ($-v$) are overlaid on the $r$ panels, while a scale bar showing a distance at the comet of 20,000~km is given on the CN panels (1 arcsec corresponds to 2134~km on October 4 and 1814~km on October 27). All four panels for a given night have the same orientation and physical scale at the comet. The October 4 image is a single frame and shows several artifacts that could not be removed without a median combination include several bright pixels in the CN frame (yellow) and a bad column in all frames (black diagonal line). On October 4, a bright star is seen near the left edge of the $r$ frames and as a diagonal (yellow) line in the upper right in CN. Two faint trailed stars are visible in the $r$ images on October 27.}
    \label{fig:image}
\end{figure}

We obtained two sets of high quality imaging: the first sets on 2019 October 4 and 27. In Figure~\ref{fig:image} we show unenhanced (top row) and enhanced (bottom row) $r$ and CN images for both October 4 (left) and 27 (right). Only a single image in each of $r$ and CN was usable on October 4 so the images suffer several artifacts and have lower signal-to-noise than the composite images on October 27. Therefore, we focus our attention in this subsection on the October 27 images though we note that the October 4 data were materially similar and yielded similar interpretations.  

\begin{figure}
\centering
	\includegraphics[width=0.72\columnwidth]{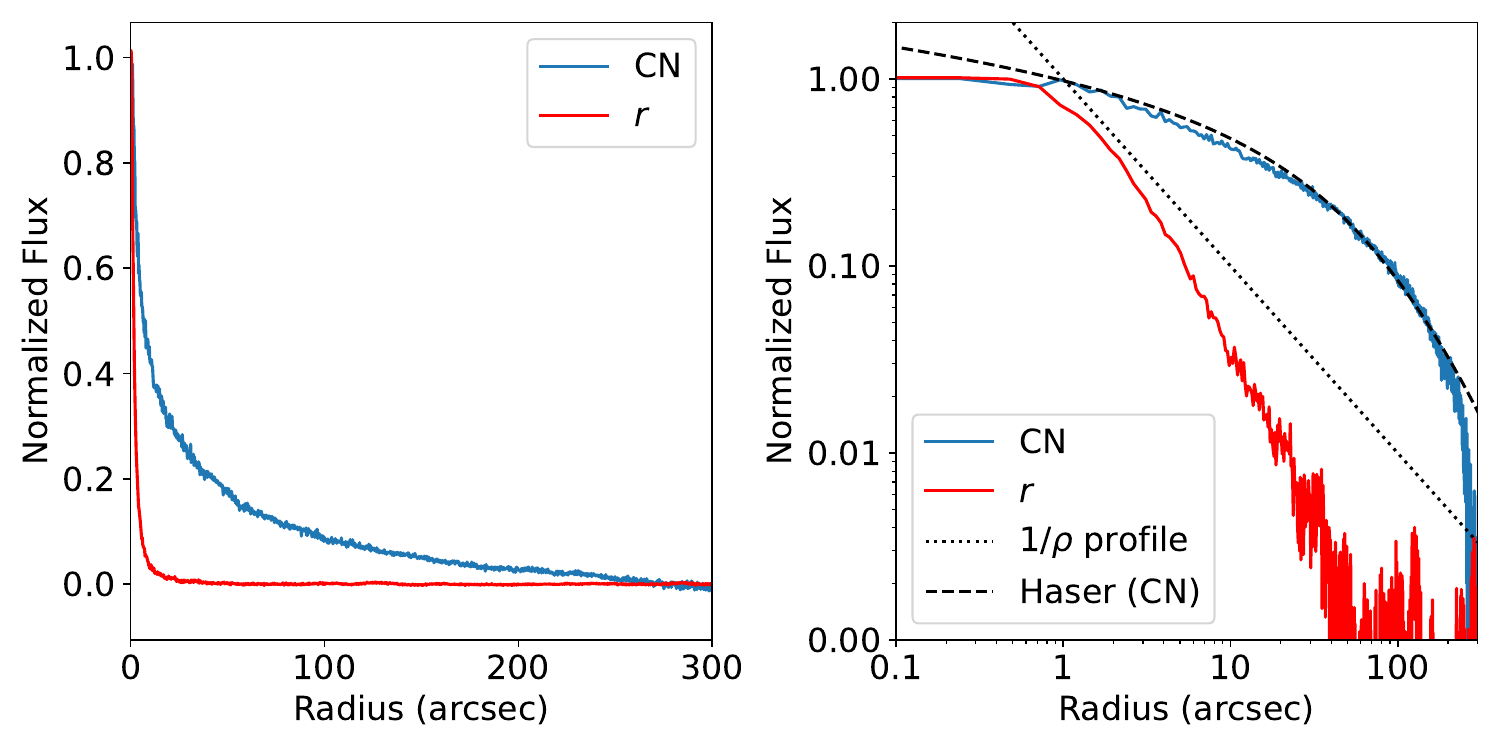}
    \caption{Radial profiles from the 2019 October 27 unenhanced $r$ and CN images shown in Figure~\ref{fig:image}. Both plots show the same data; the left plot has linear axes while the right plot has logarithmic axes and includes $1/\rho$ and Haser CN profiles overlaid to guide the eye. Both overlaid profiles have been normalized to 1 at a radius of 1 arcsec. The Haser profile was generated by the \texttt{sbpy} `Haser' routine \citep{Mommert2019} and uses the same scale lengths as described in the text (Section~\ref{sec:phot_reductions}) scaled to the date of observations. Due to the lack of absolute calibration, both the observed and model profiles are normalized. 
    }
    \label{fig:rad_prof}
\end{figure}

The $r$ image is centrally condensed with a fainter extension to the northwest at a position angle (P.A.) measured from north through east of $\sim$315$^\circ$, pointing roughly between the anti-sunward direction and the anti-velocity directions. This is commensurate with the expected tail direction, though there are hints of linearity in the enhanced structure that could plausibly be faint H$_2$O$^+$ ions since these are generally the brightest emission lines in the $r$ bandpass that would be expected to produce a linear feature \citep[cf.][]{Farnham2000,Feldman2004}. The bright spot near the center of the enhanced $r$ image surrounded by dark is presumably an artifact of the enhancement process due to a small tailward bias of the centroid, although we cannot rule out the possibility that there is a feature that we do not resolve. We saw the same feature when enhancing each individual image so it is not due to poor stacking, and stars in the field do not show a feature in this direction when the same enhancement is applied so it is not caused by the seeing. We explored a grid of centroids near our nominal solution and found that the feature disappears for centroids offset in the direction of the feature by $\sim$3 pixels. This is well within the seeing at the time of our observations ($\sim$10 pixels or $\sim$2.4 arcsec) so we do not have sufficient resolution to definitively draw a conclusion about its provenance. Beyond $\sim$2 arcsec, the slope of the $r$ radial profile is steeper than ${\rho}^{-1}$ as shown in Figure~\ref{fig:rad_prof}.

This $r$-band morphology is consistent with that reported by other authors. Intriguingly, a short dust feature was seen in {\it Hubble Space Telescope (HST)} images that may correspond to the bright spot we described. \citet{Jewitt2020} noted the ``bilobed'' appearance of the inner coma on 2019 October 12 but deferred analysis to a later paper. \citet{Manzini2020} first presented enhanced images that clearly showed the feature on October 12 and they also identified it in {\it HST} images from 2019 November 16, 2019 December 6, and 2020 January 3. \citet{Kim2020} identified ``persistent asymmetries'' in {\it HST} images on five epochs from 2019 October 12 through 2020 January 29, while \citet{Bolin2020} identified the features in four {\it HST} epochs from 2019 October 12 through 2020 January 27. The persistence of it in {\it HST} imaging over more than three months strongly suggests that it was present during our October 4 and 27 observations and may have contributed to the bright spot we noted. The feature was apparently also detected from the ground by \citet{Mazzotta2021} who noted a ``jet-like'' structure on 2019 October 19-20 and 2019 December 2.

By-eye the CN image looks symmetric. After enhancement, there is a faint asymmetry towards the northwest at a P.A. of $\sim$300$^\circ$. This is approximately co-spatial with the dust tail seen in the $r$-band image, so we conclude that this is contamination --- recall that these images were not absolutely calibrated to remove underlying dust. The enhanced images (bottom row of Figure~\ref{fig:image}) are quite noisy due to the low signal-to-noise ratio, but the excess signal is about 5-10\% of the ambient coma at a given distance. This is unusual for dust contamination in the CN bandpass, where our previous work has generally not seen evidence of dust contamination after image enhancement \citep[cf.,][]{Bair2018,Farnham2021}. We are confident, however, that the bulk of the signal in the enhanced CN images is, in fact, CN for two reasons. First, we compared the radial profile of the CN image to a Haser model (Figure~\ref{fig:rad_prof}) and found the shape to be a good match. Second, our spectrum taken immediately afterwards (Figure~\ref{fig:oct27spec}) shows strong CN emission.

\subsection{Gas and Dust Production Rates}

\subsubsection{Results from Spectroscopy and Narrowband Photometry}
\begin{figure}
\centering
	\includegraphics[width=\columnwidth]{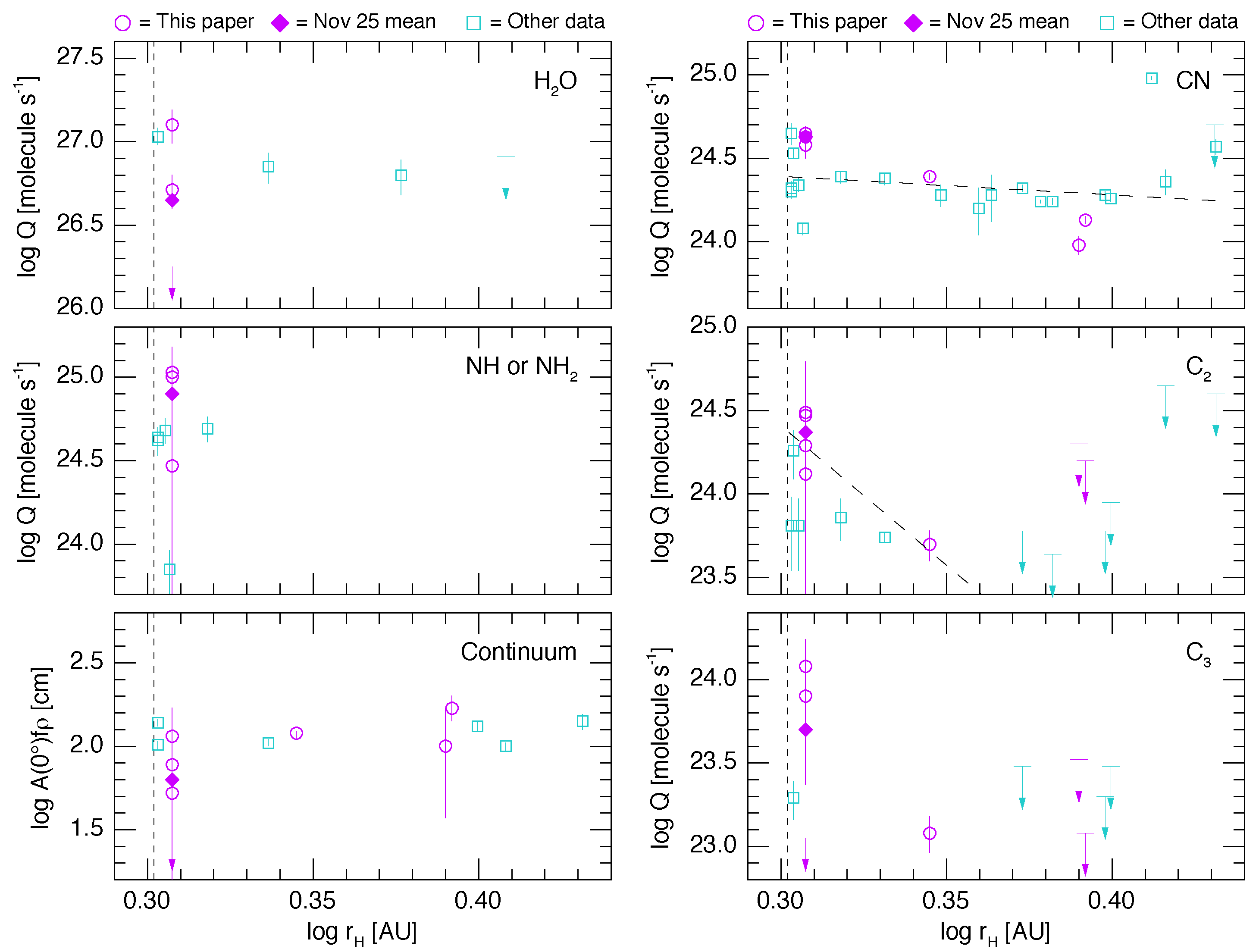}
    \caption{Logarithmic production rates for comet 2I/Borisov as a function of the log of the heliocentric distance (log $r_\mathrm{H}$). The data shown are from before its 2019 December 8 perihelion ($q$ = 2.007 au; log $q$ = 0.303 au; shown as a vertical dotted line). The data from this study are shown in purple, with individual detections represented by open circles, the average from 2019 November 25 represented by filled diamonds, and our upper limits for C$_2$ and C$_3$ denoted by dashes with attached downward arrows. Negative results or off-scale points associated with our November 25 observations are shown as purple downward arrows near the bottom of the H$_2$O, C$_3$, and continuum plots (see Table \ref{t:phot_rates}). Data from other studies are shown in light blue, with detections represented by open squares and upper limits again denoted by dashes with attached downward arrows. Our continuum values are from the green continuum while others are from either the red continuum \citep{Fitzsimmons2019, Jehin2020, Opitom2019} or V-band \citep{Xing2020}. The additional data are from \citet{Aravind2021, Deam2026, deLeon2020, Fitzsimmons2019, Jehin2020, Kareta2020, Lin2020, McKay2020, Opitom2019, Prodan2024, Xing2020}. The log $r_\mathrm{H}$-dependent fits are shown as dashed lines on the CN and C$_2$ plots only, with some points excluded from the fits, as discussed in the text. 
    }
    \label{fig:prod_rates}
\end{figure}

As previously noted, Borisov's peak total brightness reached an apparent magnitude of only $\sim15$, which is roughly the threshold at which we normally {\it begin} our compositional observations. This, combined with its limited visibility from Flagstaff, AZ and poor weather on several of our planned observing nights, means that our resulting spectroscopic and narrowband photometry datasets are sparse though we were still able to track Borisov's production rates over the course of more than seven weeks. Production rates from these nights for the gas species detected, in addition to the $A(0^{\circ})f\rho$ for dust from green continuum, are plotted in Figure~\ref{fig:prod_rates}.

On October 4 and 5, CN was confidently detected spectroscopically, but not C$_2$ or C$_3$. Our CN production rate from October 4 ($Q$(CN) = (1.36$\pm0.07)\times10^{24}$ mol s$^{-1}$) was somewhat higher than our value the following night ($Q$(CN) = (0.96$\pm0.12)\times10^{24}$ mol s$^{-1}$) at the 2-3 $\sigma$ level. While this could be partly influenced by the poorer conditions during the second night and less time on target resulting in lower signal-to-noise ratio or day-to-day variability of the comet, a more plausible reason is simply that a smaller aperture was utilized on the 5th compared to the 4th. The condition and time on target contribute to the less stringent upper limits on C$_2$ and C$_3$ production on October 5 as well (Table \ref{t:phot_rates}). While the October 4th spectrum shows higher scatter than anticipated in the vicinity of the C$_3$ feature, the strength of this putative feature does not become more convincing through the analysis steps described in Section 2.2 and thus we conclude it is an artifact. On these two nights our phase-corrected dust production rates ($A(0^{\circ})f\rho$), derived from the green continuum, are 170$\pm30$ and 100$\pm60$ cm, respectively.

Our October 27 spectrum, taken when the comet was brighter and with significantly more time on source, shows clear detections of CN, C$_2$, and C$_3$. We derived production rates for CN of (2.47$\pm0.09)\times10^{24}$ mol s$^{-1}$, and for C$_2$ and C$_3$ of (5$\pm1)\times10^{23}$~mol~s$^{-1}$ and (1.2$\pm0.3)\times10^{23}$ mol s$^{-1}$, respectively. Note that the CN emission is visibly extended even in individual frames by almost two arcminutes while the C$_2$ and C$_3$ comae appear smaller, on the order of several arcseconds in size; extractions with a range of apertures for the spectra approximate the expected Haser profiles for all species. Our green continuum $A(0^{\circ})f\rho$ value for dust on this night is 120$\pm10$ cm.

Our only successful night of observations using the photometer was 2019 November 25, approximately a month after our last spectrum and about two weeks before perihelion. We detected all five gas species that we routinely measure (CN, C$_2$, C$_3$, OH, and NH; see Table \ref{t:phot_rates}) and additionally obtained continuum measurements for deriving dust production rates. The derived production rates for each gas species, averaged from the five observational sets on this night, are (4.7$\pm$2.9) $\times$ $10^{26}$ mol s$^{-1}$ for OH (H$_2$O equivalent of 4.5 $\times$ $10^{26}$ mol s$^{-1}$); (8.5$\pm$3.6) $\times$ $10^{24}$ mol s$^{-1}$ for NH; (4.27$\pm$0.20) $\times$ $10^{24}$ mol s$^{-1}$ for CN; (2.3$\pm$0.5) $\times$ $10^{24}$ mol s$^{-1}$ for C$_2$; and (5.0$\pm$1.1) $\times$ $10^{23}$ mol s$^{-1}$ for C$_3$. For dust derived from the green continuum, the average phase-corrected production rate ($A(0^{\circ})f\rho$) is 63$\pm18$ cm. Unfortunately, the signal-to-noise is insufficient to provide a meaningful dust color. 

\subsubsection{Trends with Heliocentric Distance from Combined Studies}

In Figure ~\ref{fig:prod_rates}, we plot our gas (log $Q$) and dust (log $A(0^{\circ})f\rho$) production rates vs heliocentric distance (log $r_\mathrm{H}$), along with pre-perihelion observations from other studies obtained between 2019 September 20 and December 6. As is evident in the plots, our production rates are in reasonable agreement with those reported by others. 

The first molecule detected in Borisov was CN, in spectra from 2019 September 20 while the comet was 2.70 au from the Sun \citep{Fitzsimmons2019} (Figure~\ref{fig:prod_rates}). We obtained our first CN measurements about 14 days later ($r_\mathrm{H}$ = 2.47 au), and our production rates from October 5 and 27 are in good agreement with surrounding observations from \citet{Opitom2019} and \citet{Kareta2020}, while our last night on November 25 ($r_\mathrm{H}$ = 2.03 au) yielded production rates nearly identical to those obtained within five days of ours by \citet{Jehin2019} and \citet{Aravind2021}. Combined results from the plotted studies indicate a slight increase in CN production rates as it moved toward its early December perihelion, with a log $r_\mathrm{H}$-dependent slope of about $-$1.1 (excluding the anomalously high point at 2.58 au, as its authors acknowledge it should be taken with caution; see \citet{deLeon2020}.

Upper limits were reported for several additional species beginning in late September, but no detections of other species occurred until \citet{McKay2020} found the water tracer [O\,\textsc{i}] 6300~\AA\ line from spectra on 2019 October 11 at $r_\mathrm{H} = 2.38$~au. All reported water production rates, including ours from November 25, are in reasonable agreement given the uncertainties and the associated scatter of the data. Determining an $r_\mathrm{H}$-dependent slope for H$_2$O is problematic due to our non-detections on November 25 that lower our average production rate, but the detections from all authors suggest a slope of about -1.2 that is comparable to what was seen for CN (Figure~\ref{fig:prod_rates}). 

The first confirmed detections of C$_2$ and C$_3$ are from our spectra on October 27 (at 2.21 au; Figure~\ref{fig:prod_rates}), and they are in agreement with most production rates obtained by others thereafter \citep{Aravind2021, Lin2020}; the exception is the C$_2$ production rates reported by \cite{Deam2026} which are systematically lower and likely the result of an aperture effect due to their chosen aperture sizes for spectrum extraction (C. Opitom, personal communication, Dec 2025). Production rates for both C$_2$ and C$_3$ increased much more rapidly than CN; over the small range when they were detected, the log $r_\mathrm{H}$-dependent slope for C$_2$ is $-$17 (excluding the systematically lower \cite{Deam2026} data, but note it is still quite steep at -12 if these points are included). Determining a slope for C$_3$ is problematic due to it being a weaker feature (and its resulting non-detections in half of our photometry sets in late November), though it appears to be similarly steep. 

Of the species we measure, products of ammonia (as NH for us and NH$_2$ from others) were the last to be definitively detected in Borisov, with all pre-perihelion measurements occurring within a month of perihelion. Our NH data from November 25 are bracketed by observations from \cite{Deam2026}, who measured NH$_2$ on several nights from 2019 November 14 (2.08 au) to December 6. The data are in reasonable agreement given the uncertainties and imply ammonia production rates did not change significantly over this short interval. An additional measurement from \cite{Prodan2024} on November 26 gave an NH$_2$ production rate several times lower than both ours from November 25 and that from \cite{Deam2026} on November 26.

Borisov's behavior with heliocentric distance is unique in that it has such a large discrepancy between the behavior of CN and the carbon-chain species. For {\it Q}(CN), which we detected during each of our observing attempts, our data indicate a steady increase in pre-perihelion production rates of about 70$\%$. For both {\it Q}(C$_2$) and {\it Q}(C$_3$) the increase is much steeper, though we detected these species only in late October and late November; over this distance (just 2.213 to 2.028 au) our production rates increased by over 300$\%$. 

The overall increase in CN and water production rates is somewhat less than a heliocentric distance dependence of $r_\mathrm{H}^{-2}$. This suggests there might have been some hyperactivity at larger distances similar to what we have seen in several dynamically new Oort Cloud comets, which are often hypothesized to be active over most, if not all, of their illuminated surface and are usually observed to exhibit a shallow increase in activity \citep[e.g.,][]{Whipple1978}. For C$_2$ and C$_3$, however, its behavior mimics the short-period comets \citep[cf.][]{Schleicher2003,Knight2012}, which often have discrete source regions on their nuclei that rapidly turn on due to the changing illumination from the Sun at certain points in their orbits and results in steeper $r_\mathrm{H}$-dependencies like what was seen for these two species in Borisov. We emphasize that while a rapid increase in production rates is not uncommon, the increase for C$_2$ and C$_3$ in Borisov is among the steepest we have observed, and the large discrepancy between the rate of increase for CN and that of C$_2$ and C$_3$ is extremely unusual. 

Our combined dust production rates (log $A$(0$^\circ$)$f\rho$) from broadband $R$ imaging (Fig.~\ref{fig:robo-phot}) and spectroscopy and photometry observations in the green continuum (Fig.~\ref{fig:prod_rates}) indicate Borisov's peak dust production of 130 cm occurred about 50 days before perihelion, around 2019 October 19 at approximately 2.3 au and between our early and late October spectroscopy observations. Dust production then steadily decreased as the comet approached perihelion with a log $r_\mathrm{H}$-dependent slope of about 1.0. The $A$(0$^\circ$)$f\rho$ measurements from others are in good agreement with ours as can be seen in Fig.~\ref{fig:prod_rates}, and \citet{Jehin2020} confirm this early peak in dust production (note that we do not show the $r_\mathrm{H}$-dependent fit of the dust production rates in Fig.~\ref{fig:prod_rates} due to its peak and subsequent decline before even reaching its closest approach to the Sun). Our only post-perihelion data point is from the $R$ imaging eight days after perihelion on 2019 December 16 (Fig.~\ref{fig:robo-phot}), and indicates dust production continued to drop after the comet's perihelion passage, also in agreement with \citet{Jehin2020}. 

All of our gas production rates are from before perihelion; as such, we did not directly measure changes in gas production after Borisov completed its closest approach to the Sun. The results combined from multiple studies, however, suggest CN, C$_2$, C$_3$ and NH$_2$ production remained steady through late December \citep{Aravind2021, Deam2026}. CN then steadily decreased from the end of December through the end of February, before Borisov's early March outburst and subsequent fragmentation \citep{Jewitt2020}, while C$_2$ and NH$_2$ dropped more quickly \citep{Deam2026}. Its CO production increased through perihelion and continued to rise until at least a month after \citep{Cordiner2020, Yang2021}, but water peaked near perihelion \citep{Xing2020, Bodewits2020}. Dust reached its peak production about 50 days before perihelion, and both water and dust were observed to decrease rapidly following Borisov's closest approach to the Sun.

\subsubsection{Nucleus Properties}

We propose that the differing behaviors observed between the species, with different species peaking significantly before, at, or after perihelion, imply either compositional heterogeneity between source regions exposed before and after perihelion, or the exposure of fresh subsurface materials with different chemical abundances, e.g., as the result of surface erosion of Borisov's nucleus or the propagation of thermal energy into the interior. Since its composition was measured to change before the comet even reached perihelion, with different species exhibiting different behaviors, we do not think seasonal changes in source regions are the main culprit. Instead, we speculate that a significant alteration of Borisov's outer layer occurred during its interstellar journey, and the sublimation of this crust as it approached the Sun produced the discrepancies that we and others observed. This idea is supported by analyses from \citet{Kim2020}, who estimate that $\sim$0.4 m of crust eroded between August 2019 and January 2020, and suggest that the ejected materials observed before and during Borisov's perihelion passage may not necessarily have been pristine. Our images showing Borisov with a nearly symmetric coma imply activity over its entire surface also support the idea that there was sublimation of an altered layer that covered its nucleus; if there were discrete source regions with different compositions, we would expect to see a discernible asymmetry in the comet's coma that would reflect these isolated source areas. While \cite{Yang2021} report that the rate at which fresh subsurface materials were unveiled on Borisov's nucleus was not compatible with its fast decline in post-perihelion water production rates, we argue that the erosion of near-surface layers with disparate compositions is entirely plausible, and is strengthened when considering that the different species increased and decreased at differing rates and heliocentric distances.

Borisov was very faint, and its production rates are correspondingly low when compared to most Oort cloud comets that we have observed at similar heliocentric distances. Assuming that the active fractions of Borisov and dynamically new comets are similar, the discrepancy in brightness for Borisov implies it has a smaller nucleus than most dynamically new and young long-period Oort Cloud comets that we have observed. We can estimate the active area necessary to support our measured $Q$(OH) by using the \citet{Cowan1979} sublimation rates provided by the Planetary Data System's Small Bodies Node\footnote{\href{https://pdssbn.astro.umd.edu/tools/ma-evap/index.shtml}{https://pdssbn.astro.umd.edu/tools/ma-evap/index.shtml}}. Converting $Q$(OH) to $Q$(H$_2$O) by $1.361~r_\mathrm{H}^{-0.5}~Q$(OH) \citep{Cochran1993, Schleicher1998}, this yields an active area of ${\sim}4{\times}10^5$~m$^2$ for the subsolar case and ${\sim}4.5{\times}10^6$~m$^2$ for the isothermal case, representing the theoretical highest and lowest possible active areas. Assuming the nucleus was active over its entire surface, these correspond to effective radii ($r_\mathrm{n}$) between 180~m and 600~m, consistent with what has been inferred by \cite{Jewitt2020} ($r_\mathrm{n}$ $\leq 0.5$ km), \cite{Xing2020} ($r_\mathrm{n}$ $\geq 0.37$ km), and \cite{Ye2020} ($r_\mathrm{n}$ $\leq 0.4$ km).

\subsection{Chemical Composition}

\subsubsection{Compositional Changes with Heliocentric Distance}

\begin{figure}
\centering
	\includegraphics[width=0.5\columnwidth]{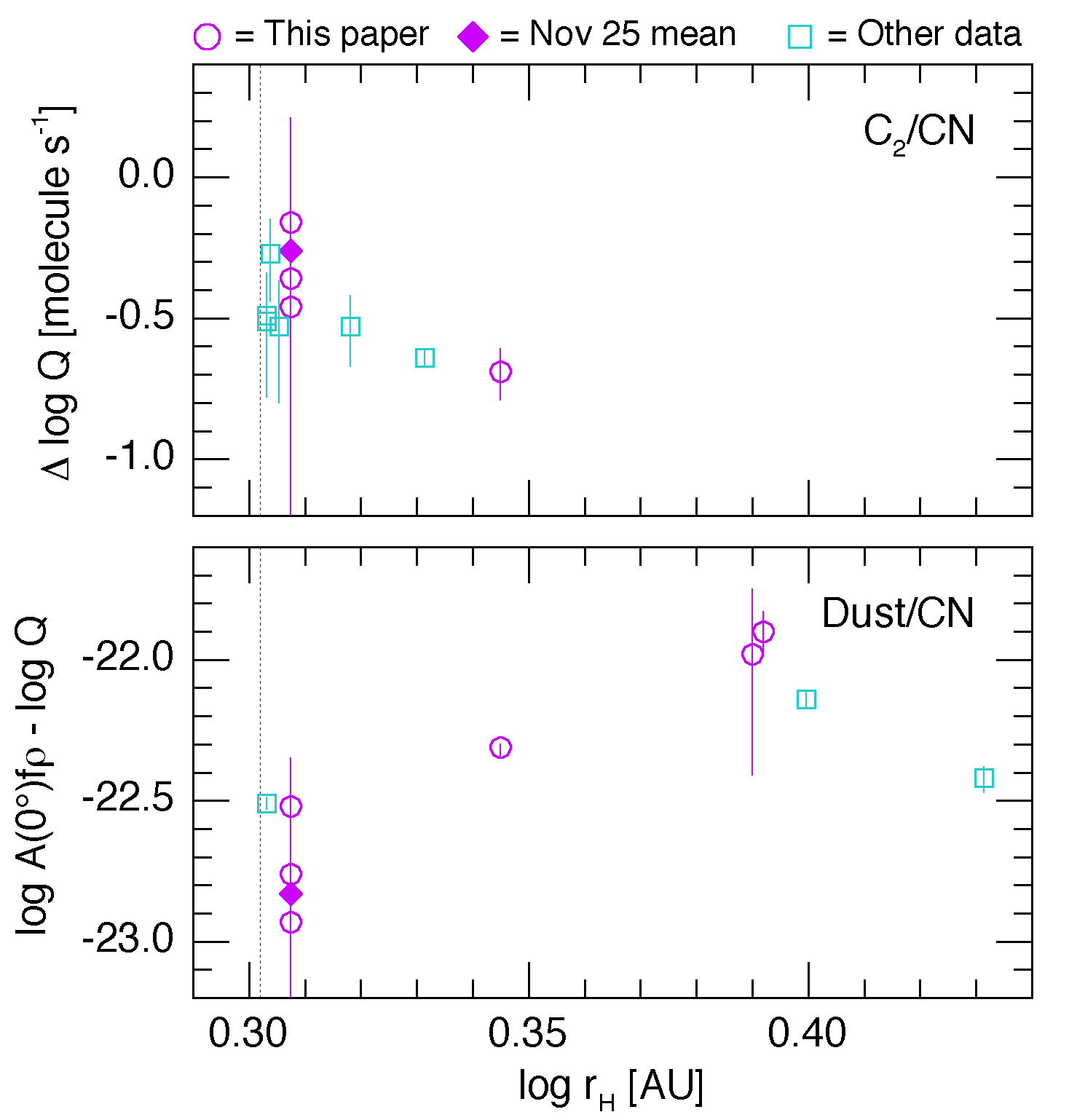}
    \caption{Pre-perihelion logarithmic production rate ratios for comet 2I/Borisov as a function of the log of the heliocentric distance (log $r_\mathrm{H}$). Colors and symbols are the same as in Figure ~\ref{fig:prod_rates}. Borisov's C$_2$-to-CN production rate ratios greatly increased as the comet approached perihelion, while its dust production rate ratios as $A$(0$^\circ$)$f\rho$-to-CN peaked significantly before perihelion. The additional data are from \citet{Aravind2021, Deam2026, Fitzsimmons2019, Jehin2020, Lin2020, Opitom2019, Xing2020}. 
    }
    \label{fig:ratio_logr}
\end{figure}

As previously discussed, the production rates measured for Borisov revealed its various molecular species and dust had disparate behaviors with heliocentric distance. It is therefore not surprising that most of the resulting ratios of its production rates (i.e. abundance ratios) also changed with heliocentric distance. In our dataset the largest change occurred for the compositional abundance ratios of [log {\it Q}(C$_2$) $-$ log {\it Q}(CN)] (hereafter C$_2$-to-CN), and [log {\it Q}(C$_3$) $-$ log {\it Q}(CN)] (C$_3$-to-CN), both of which increased by about 150$\%$ during the month between our observing runs where all three of these species were detected. This is a direct result of the rapidly increasing production rates for both C$_2$ and C$_3$ and the slower rise for CN over this small range of heliocentric distance (just 2.21 to 2.03 au); we have plotted this for C$_2$-to-CN in the top panel of Fig.~\ref{fig:ratio_logr} for all studies where this ratio is available. Specifically, between October 27 and November 25 our C$_2$-to-CN abundance ratios increased from $-$0.7 to $-$0.3 while C$_3$-to-CN increased from $-$1.3 to $-$0.9. Measurements from \citet{Lin2020} on November 4 ($r_\mathrm{H}$ = 2.15 au) and \citet{Aravind2021} on November 30  ($r_\mathrm{H}$ = 2.01 au) also indicate an increase in C$_2$-to-CN from $-$0.7 to $-$0.27, respectively, and support our observed shift in composition. \citet{Aravind2021} additionally obtained C$_3$-to-CN on November 30; their value of $-$1.23 is lower than ours on November 25 but is comparable within the uncertainties. The C$_2$-to-CN values reported by \cite{Deam2026} from November 14 (2.08 au) to December 6 (2.01 au) are a factor of 1.5 to 2 lower than those reported by us and others, but this can likely be explained by their chosen C$_2$ aperture sizes for extractions, as noted above in Section 3.2.2. We note that Borisov's [log {\it Q}(C$_3$) $-$ log {\it Q}(C$_2$)] (C$_3$-to-C$_2$) ratio did not demonstrably change between our observations on October 27 and November 25, where the ratios were $-$0.6 and $-$0.7, respectively, as both species were increasing at similar rates.

Since we only detected OH and NH during our last night of observations, we cannot discern if there are any trends with heliocentric distance for abundance ratios involving these two species using only our data. The combination of multiple studies, however, suggests that [log {\it Q}(CN) $-$ log {\it Q}(H$_2$O)] (CN-to-H$_2$O) abundance ratios remained more or less steady from mid-October until perihelion, and that ammonia as either NH or NH$_2$-to-H$_2$O decreased very slightly, reflecting that ammonia production appears to have peaked before or during mid-November, while H$_2$O production rates continued to rise somewhat until perihelion. 

The dust-to-gas abundance ratios as [log $A$(0$^\circ$)$f\rho$ $-$ log {\it Q}(CN)] (log $A$(0$^\circ$)$f\rho$-to-CN) approximately follow the same trend as the $A$(0$^\circ$)$f\rho$ production rates seen in Figure~\ref{fig:robo-phot}, with the abundance ratio peaking between 7 and 9 weeks before perihelion (between 2.47 and 2.21 au). The following decrease for the abundance ratios is larger, however, since CN production was continuing to rise as Borisov approached perihelion while dust production dropped (Figure~\ref{fig:ratio_logr}, lower panel). The highest log $A$(0$^\circ$)$f\rho$-to-CN value we measured was $-$21.90 on 2019 October 4, 65 days before perihelion at 2.47 au; this ratio then dropped to $-$22.83 by 13 days prior to perihelion on November 25 at 2.21~au. The combined dust abundance ratios reported by \citet{Fitzsimmons2019}, \citet{Opitom2019}, and \citet{Jehin2019} support this trend.

The compositional changes with heliocentric distance seen in Borisov are extremely unusual, and no other comet in the Lowell Observatory database has exhibited such large variations in abundance ratios during its perihelion passage. Note that we have documented a systematic trend of decreasing C$_2$-to-CN abundance ratios with increasing heliocentric distance among the many comets in the Lowell database, but the change is much smaller than what was seen for Borisov, and heliocentric trends are absent or minimal in the abundance ratios for the remaining species that we measure, including for C$_3$-to-CN (see \citealt{Bair2025}).

\subsubsection{Compositional Taxonomy}

\begin{figure}
\centering
	\includegraphics[width=0.95\columnwidth]{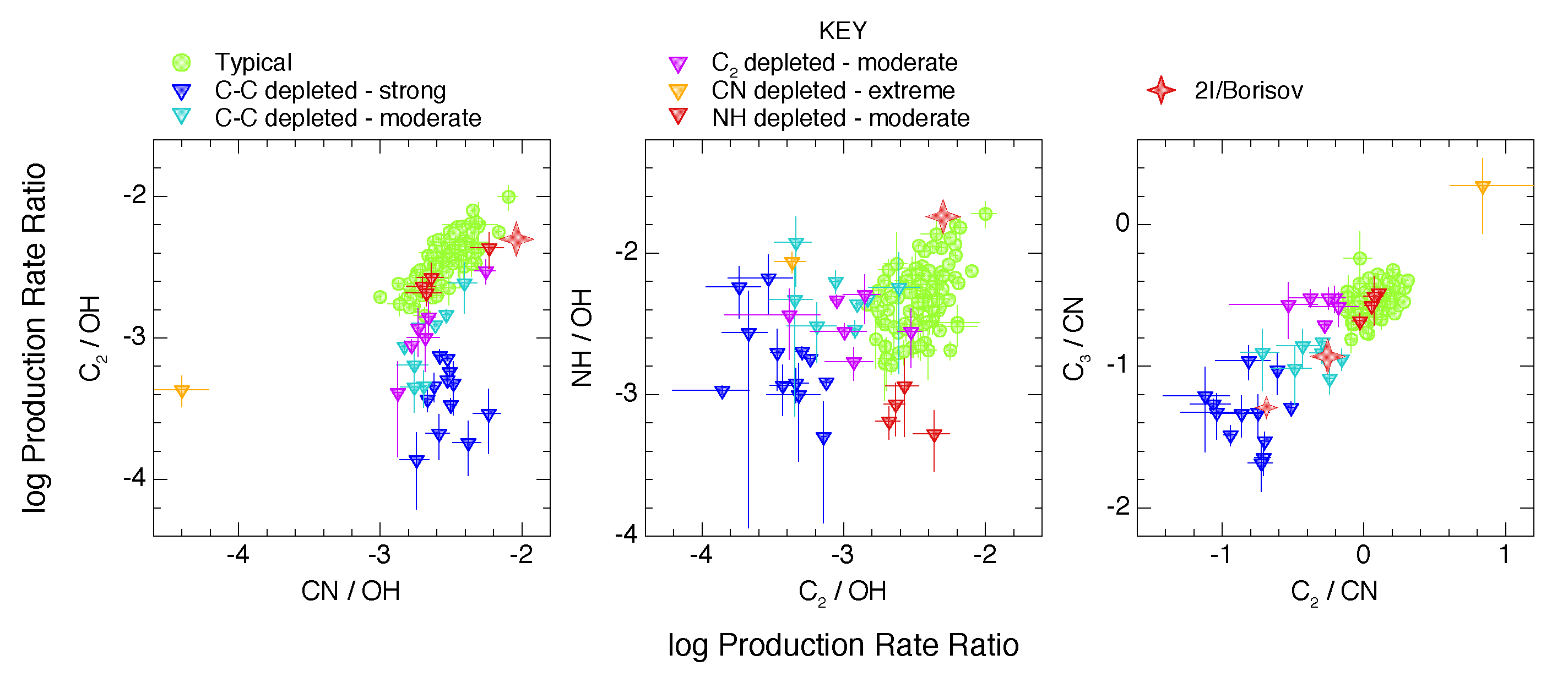}
    \caption{Logarithmic abundance ratio plots from the restricted subset of our database, with the addition of 2I/Borisov as a red diamond star. Borisov's mean abundance ratios from 2019 November 25 are represented by the larger diamond star and appear in all 3 plots. Our data from October 27 are represented by a smaller symbol and only appear in the last plot as we did not observe OH and NH on this night; its C$_2$-to-CN and C$_3$-to-CN abundance ratios changed greatly between the two nights, moving Borisov from the strongly carbon-chain depleted class to the moderately carbon-chain depleted class as it approached perihelion.}
    \label{fig:tax}
\end{figure}

To meaningfully compare Borisov's abundance ratios to solar system comets, we utilize the restricted subset of 133 comets from Lowell Observatory's narrowband photometry database and show the taxonomy plots for key abundance ratios in Figure~\ref{fig:tax}. In brief, this restricted subset consists of 133 comets, with observations spanning nearly 50 years, for which we consider their production rates to be well-determined: they have a minimum of three measurement sets, two of which must include detections of all five of the gas species we measure, and their observations were obtained over at least two nights at heliocentric distances of less than 3 au. Additional details and a more in-depth explanation of the compositional groupings can be found in \citet{Bair2025}.

For Borisov, our abundance ratios from November 25 (the only night we measured OH and NH) indicate that its production rate ratios of all gas species with respect to OH place it within the typical compositional class. Its mean [log {\it Q}(CN) $-$ log {\it Q}(OH)] (CN-to-OH) ratio of $-$2.0, however is higher than any comet in the restricted subset, as can be seen in the left-hand panel of Figure~\ref{fig:tax} where Borisov is represented by a large red diamond star; the two comets with similarly high CN-to-OH ratios are C/2019 Y1 (ATLAS) and C/2020 F3 (NEOWISE), both of which are dynamically old long-period comets. Its mean [log {\it Q}(NH) $-$ log {\it Q}(OH)] (NH-to-OH) ratio of $-$1.7 is also among the highest we have measured (middle panel), and the only comet on this plot with higher NH-to-OH is C/2019 Y1 (ATLAS). We note that 1P/Halley, C/1995 O1 (Hale-Bopp), and C/1996 B2 (Hyakutake) also have high NH-to-OH ratios and are in the vicinity of Borisov on this plot. All of our other abundance ratios for Borisov with respect to OH, as well as [log {\it Q}(NH) $-$ log {\it Q}(CN)] (NH-to-CN), are somewhat high but not unusually so and are among those seen for the typical comets (i.e.\ those exhibiting no anomalous compositions, green circles on the plots). We note that \cite{Deam2026} found their NH$_2$-to-CN values to be high compared to most comets from other studies, and they suggest a relatively rich NH$_2$ composition.

Borisov's C$_2$-to-CN and C$_3$-to-CN ratios, conversely, are most similar to the comets in our carbon-chain depleted classes, but unlike any other comet we have observed, the amount of depletion changed in Borisov with heliocentric distance (Figure~\ref{fig:tax}, right panel). Our earliest compositional measurements on October 27 (the smaller red diamond star) yielded log C$_2$-to-CN and log C$_3$-to-CN ratios of $-$0.7 and $-$1.3, respectively, which placed Borisov in the strongly carbon-chain depleted class. By November 25 (represented by the large red diamond star), however, our values for log C$_2$-to-CN and log C$_3$-to-CN increased to $-$0.3 and $-$0.9, respectively, with a resulting large shift in its compositional class to becoming just moderately carbon-chain depleted. We emphasize that the strongly-carbon chain depleted comets in our database, many of which were observed over similar heliocentric distances, all remain in the strongly depleted class over the course of their observations, as detailed in \citet{Bair2025}.

Finally, for its mean dust-to-gas ratios of both log $A$(0$^\circ$)$f\rho$-to-CN and log $A$(0$^\circ$)$f\rho$-to-OH ($-$22.2, and $-$24.9, respectively) we find that Borisov's values are near the middle of the values that we have observed for other comets. While $A$(0$^\circ$)$f\rho$-to-CN changed with heliocentric distance (and presumably also $A$(0$^\circ$)$f\rho$-to-OH since water was observed to increase at a similar rate as CN), we do not find a correlation of dust-to-gas ratios and compositional classes in our database. We note the change here, but it had no affect on its taxonomic classification.

\subsection{Placing Interstellar Comets into Context}

While the known interstellar comets share similarities with those originating in our solar system, there are key differences in their orbits and evolutionary histories. Interstellar comets have significantly higher velocities than solar system comets, and are therefore not gravitationally bound to the Sun. Since they are traveling faster than Oort Cloud comets, an interstellar comet's nucleus hasn't received as much solar insolation while moving between any two heliocentric distances which may result in delayed production of volatiles relative to those from the Oort Cloud. There is additionally the possibility that these objects underwent evolution in their home solar systems before being ejected, and, depending on how long ago they were ejected, they may have a more or less irradiated crust than the Oort Cloud comets. The removal of this outer layer upon approaching the Sun could produce changes in chemical composition with heliocentric distance like what was seen in Borisov, as a more pristine interior becomes exposed. 

With the more recent discovery of 3I/ATLAS (C/2025 N1) on 2025 July 1, we now have a second interstellar comet to which we can compare the findings of our study and those of others; we will address this in a forthcoming paper of our narrowband photometry results for 3I/ATLAS. We emphasize the importance of large, long-term surveys of comets, such as ours at Lowell Observatory \citep{AHearn1995, Schleicher2016, Bair2025}, in addition to those from \citet{Fink2009}, \citet{Crovisier2009}, \citet{Cochran2012}, \citet{DelloRusso2016}, and \citet{Lippi2020}, among others, that are critical for eliciting how similar or different these interstellar comets are as compared to those originating in our solar system. Furthermore, observations are rarely made of comparably faint and/or distant objects such as 2I/Borisov and 3I/ATLAS, making direct comparisons of some aspects of their composition and evolution difficult.

\section{Summary and Conclusions} 

We report on pre-perihelion spectroscopy, photometry, and imaging of Comet 2I/Borisov from 2019 September to December.  Our broadband photometry showed subtle ($<$20\%) variations in \afrho[0\degr] (a dust production rate proxy) with heliocentric distance, peaking about 50~days before perihelion, and decreasing thereafter.  Our first detection of CN was on October 4, and we continued to detect it on each of our subsequent observing nights through November 25. We first detected both C$_2$ and C$_3$ on October 27, and these are the earliest reported detections of these two species. During our final observing night on November 25, we detected all five of the gas species that we routinely measure in comets $-$ OH, NH, CN, C$_3$, and C$_2$ $-$ as well as dust from three continuum points. Our measurements show an increase of about 70$\%$ for CN production rates as Borisov moved from 2.47 to 2.03 au, similar to many dynamically new and long period solar system comets over comparable heliocentric distances. For C$_2$ and C$_3$ our observational timeline is shorter (2.21 to 2.03 au), and for both we observed a rapid increase in production rates of over 300$\%$. An increase of this extent is not common as compared to other comets in our narrowband photometry database, though we have previously seen a similarly rapid increase in a few short-period Jupiter-family comets (e.g., \citealt{Knight2012}) known to be caused by seasonal variations associated with localized active regions.

Multiple studies, including this one, confirm that Borisov contains all of the same constituents that we regularly observe in solar system comets, indicating a similarity in formation processes across planetary systems. Borisov is unique in the context of our database, however, in that its production rates for CN and the two carbon-chain species increased at extremely different rates as the comet approached perihelion between 2.21 to 2.03 au; this is highly unusual, because in other comets we have observed all three of these carbon-bearing species increase at similar rates to one another over these same heliocentric distances. This discrepancy resulted in a change of compositional classes as Borisov moved closer to the Sun, from strongly carbon-chain depleted on October 27 (log C$_2$-to-CN ratio of $-$0.7) to just moderately carbon-chain depleted on November 25 (log C$_2$-to-CN ratio of $-$0.3) $-$ a phenomenon we have not observed in solar system comets \citep{Bair2025}. Additional studies indicate that the production rates of other species and dust also changed by different amounts and peaked at disparate heliocentric distances, resulting in further compositional anomalies as Borisov approached and receded from the Sun. Borisov's CN-to-OH and NH-to-OH production rate ratios are among the highest we have measured, while its abundance ratios for the carbon-chain species relative to OH are within the range for typical comets in our database, as is its NH-to-CN ratio. 

We hypothesize that the compositional changes seen in Borisov as it approached the Sun are likely the result of the exposure of subsurface materials due to surface erosion or the propagation of thermal energy into the interior, as the Sun ablated an outer layer or crust that had been altered during Borisov's tens of millions $-$ or even billions $-$ of years of interstellar travel \citep[cf.][]{Lintott2022, Hopkins2025}. This is supported by the different heliocentric distances at which its measured molecular species peaked, with some peaking week(s) before perihelion, some at perihelion, and others significantly after. Our images showing a nearly symmetric coma suggest Borisov was active over its illuminated surface and support this conclusion, since the alternate potential cause, compositional heterogeneity between source regions, would likely cause more obvious coma asymmetries. In conclusion, a combination of studies have revealed that while 2I/Borisov appears to contain the same chemical components as the comets originating within our Solar System, its relative abundances and behaviors with heliocentric distance are highly unusual, reflecting its interstellar origin and unique journey.

\begin{acknowledgments}
This work was initially supported by NASA's Solar System Observations Program grant 80NSSC18K0856. We also gratefully acknowledge support from the Marcus Cometary Research Fund at Lowell Observatory for the completion of these analyses. We thank Brian Skiff for assistance with the Lowell 31-in telescope scheduling, and, finally, we thank the two anonymous referees whose suggestions improved the quality of this article.

The views expressed in this article are those of the authors and do not reflect the official policy or position of the U.S. Naval Academy, Department of the Navy, the Department of Defense, or the U.S. Government.
\end{acknowledgments}

\begin{contribution}

ANB led paper writing and analyses of narrowband photometry and compositional assessment.
DGS acquired the narrowband photometry, and co-led analyses of narrowband photometry and compositional assessment.
TK led spectroscopy analyses. 
MMK acquired all LDT data and led narrowband imaging analyses.
MSPK acquired LDT data on October 4-5 and led the analyses for the 31-in imaging. 
All authors contributed to writing and editing the text.

\end{contribution}

\facilities{Lowell Discovery Telescope (LDT; 4.2 m); Lowell Observatory 42 inch (1.1 m) John S. Hall Telescope; Lowell Observatory 31 inch (0.8 m) telescope}
\software{
    \texttt{Astropy} \citep{astropy2013,astropy2018,astropy2022}, 
    \texttt{Calviacat} \citep{Kelley2019-calviacat}, 
    \texttt{IDL} \citep{idl_software}, 
    \texttt{NumPy} \citep{vanderwalt2011,harris2020}, 
    \texttt{Pypeit}  \citep{2020JOSS....5.2308P}, 
    \texttt{sbpy} \citep{Mommert2019}, 
    \texttt{SciPy} \citep{virtanen2020}
}

\bibliography{refs}
\bibliographystyle{aasjournalv7}

\end{document}